\documentclass[useAMS,usenatbib]{mn2e}

\usepackage{graphicx}
\usepackage{subfigure}
\usepackage{amsmath}
\usepackage{amssymb}

\newcommand{\brunt}{Brunt-V\"ais\"al\"a \,}
\def\gsim{\;\rlap{\lower 2.5pt
 \hbox{$\sim$}}\raise 1.5pt\hbox{$>$}\;}
\def\lsim{\;\rlap{\lower 2.5pt
   \hbox{$\sim$}}\raise 1.5pt\hbox{$<$}\;}

\begin{document}
\bibliographystyle{mn2e}

\title[Galaxy Motions, Turbulence and Conduction in Clusters of Galaxies]{Galaxy Motions, Turbulence and Conduction in Clusters of Galaxies}
\author[M. Ruszkowski \& S.P. Oh]{M. Ruszkowski$^{1,2}$ and S. Peng Oh$^{3}$ \\
$^{1}$Department of Astronomy, University of Michigan, 500 Church Street, Ann Arbor, MI 48109, USA; 
e-mail: mateuszr@umich.edu (MR) \\
$^{1}$The Michigan Center for Theoretical Physics, 3444 Randall Lab, 450 Church St, Ann Arbor, MI 48109, USA \\
$^{2}$Department of Physics, University of California, Santa Barbara, CA 93106, USA; 
e-mail: peng@physics.ucsb.edu (SPO)}
\maketitle

\begin{abstract}

Unopposed radiative cooling in clusters of galaxies results in excessive mass deposition rates onto the central brightest cluster galaxy.
However, the cool cores of galaxy clusters are continuously heated by thermal conduction and turbulent heat diffusion due to 
minor mergers or the galaxies orbiting the cluster center. 
These process can either reduce the energy requirements for AGN heating of cool cores, or they can prevent overcooling altogether. 
We perform 3D MHD simulations including field-aligned thermal conduction and self-gravitating 
particles to model this in detail. Turbulence is not confined to the wakes of galaxies but is instead volume-filling, due to the excitation of large-scale g-modes. 
We systematically probe the parameter space of galaxy masses and numbers to assess when the cooling catastrophe is prevented.
For a wide range of observationally motivated galaxy parameters, we find that the magnetic field is randomized by stirring motions, restoring the conductive heat flow to the core. The cooling catastrophe either does not occur 
or it is sufficiently delayed to allow the cluster to experience a major merger that could reset the conditions in the intracluster medium. Whilst dissipation of turbulent motions (and hence dynamical friction heating) is negligible as a heat source, turbulent heat diffusion is extremely important; it predominates in the cluster center. However, thermal conduction becomes important at larger radii, and simulations without thermal conduction suffer a cooling catastrophe. Conduction is important both as a heat source and to reduce stabilizing buoyancy forces, enabling more efficient diffusion. Turbulence enables conduction, and conduction enables turbulence. 
In these simulations, the gas vorticity---which is a good indicator of trapped g-modes--increases with time. The vorticity growth is approximately mirrored by the 
growth of the magnetic field, which is amplified by turbulence. 
\end{abstract}


\section{Introduction}
The intracluster medium (ICM) in many galaxy clusters has central cooling times shorter then the Hubble time.
Radiative cooling should lead to large accumulation of cold material 
in their centers; however, there is no observational evidence for such gas.
This can be understood if some source of heating balances cooling in the ICM.
The heating mechanisms invoked to explain this overcooling problem 
involve AGN ``radio mode'' heating (e.g., \citet{binney95,churazov02,fabian03,ruszkowski04,ruszkowski04a,scannapieco08}), 
preheating by AGN \citep{mccarthy08}, cosmic rays from AGN \citep{guo08a,sharma09}, 
supernovae, turbulent mixing \citep{kim03a,voigt04,dennis05},
thermal conduction \citep{zakamska03,kim03}, a combination of thermal conduction and AGN \citep{ruszkowski02} 
and dynamical friction \citep{elzant04,kim05,kim07}; see \citet{conroy08} and 
\citet{mcnamara07} and references therein for reviews of the above mechanisms).\\
\indent
Conduction alone is unlikely to offer the complete solution to the overcooling problem for the full range of
cluster masses, as its strong temperature dependence implies that it is less effective in lower mass clusters. Furthermore, thermal conduction is well known to be an unstable heating mechanism, either failing to avert a cooling catastrophe, or leading to an isothermal temperature profile \citep{bregman88,guo08a,conroy08}. 
Nevertheless, thermal conduction may entirely suppress cooling in non cool-core (NCC) clusters and reduce 
the constraints on the required energy injection by AGN in cool-core (CC) clusters \citep{guo08b}; indeed, it is different to stabilize massive clusters with AGN feedback alone \citep{conroy08}, and a second heat source (such as conduction) is generally required. Besides offsetting radiative losses and stemming a cooling catastrophe, conduction can have important implications for establishing the observed bimodality in cluster core entropy \citep{guo08b}, and the star formation threshold in brightest cluster galaxies \citep{voit08}. Indeed, a sudden increase in conduction (due to say, turbulence from an AGN outburst or a merger) could mediate a cool core to non cool-core transition \citep{guo09}. 

Thermal conduction can be strongly suppressed by magnetic fields that are known to be present in the ICM (e.g., En{\ss}lin et al. (2003), 
Vogt et al. 2003). However, interest in thermal conduction as a potential heating mechanism was revived by 
\citet{narayan01} who suggested that even in the presence of tangled B-fields, the level of conduction can be 
an appreciable fraction of the Spitzer-Braginskii value.
Computing the exact magnitude and distribution of the effective 
conductivity of the ICM is further complicated by buoyancy instabilities which re-orient the magnetic field. When temperature increases in the direction of gravity, as in the cluster outskirts, the magnetothermal instability (MTI; \citet{balbus00,parrish05}) tends to make the B-fields radial, thereby increasing the effective conduction. On the other hand, in cool cores where temperature decreases in the direction of gravity, the heat flux buoyancy instability (HBI; \citet{quataert08,parrish09,bogdanovic09}) 
tends to reorient the fields in the 
direction perpendicular to that of gravity, effectively shutting down thermal conduction.

However, these instabilities do not operate in a static atmosphere. {\it Chandra} and XMM observations show that the cluster gas is rarely in perfect hydrostatic equilibrium. 
Sloshing motions due to minor mergers, AGN, or galaxy motions can 
continuously and significantly perturb the gas, as has been repeatedly seen in many disparate numerical simulations \citep{evrard90,norman99,nagai03,vazza09,vazza10,ascasibar06,zuhone10}. Current observational evidence for turbulence ranges from the analysis of pressure maps \citep{schuecker04}, its effect on resonant-line scattering \citep{churazov04}, and Faraday rotation maps \citep{vogt05,enslin06}, as well as constraints on turbulent line widths \citep{sanders09}. 
Low levels of turbulence in the ICM can randomize the field configuration set up by the HBI 
and restore the heat flow to the core \citep{ruszkowski10,parrish10}. 
Both of these works modeled the ICM turbulence via a simple driving mechanism to determine the level of turbulence required
to effectively restore thermal conduction. This approach did not allow us to link the level of turbulence to the physical properties
of the cluster (such as mechanical luminosity of the central AGN or the properties of cluster galaxies). Also,
the driving mechanism led, by construction, to volume-filling turbulence which was very 
effective in randomizing the magnetic field. While the low amplitude of the required subsonic turbulence is eminently feasible ($v_{t} \sim 150 {\rm km \, s^{-1}} \sim 0.1 c_{s}$), the realism of volume-filling turbulence is less clear. For instance, both analytic calculations \citep{subramanian06} and numerical simulations \citep{iapichino08} predict that turbulence due to galaxy wakes should not be volume-filling ($f_{\rm V} \lsim 0.2-0.3$), as turbulence is largely confined to 'streaks' behind orbiting galaxies. 

In this paper, we extend our previous work and perform three-dimensional MHD simulations of the effect 
of turbulence driven by galaxy motions on the properties of the anisotropic thermal conduction.
We show how the trapping of gravity modes excited by the orbiting galaxies can lead to volume-filling turbulence of the right 
magnitude to restore conductive heat flow. We demonstrate how these subsonic motions generate vorticity and lead to the 
growth of magnetic field via kinematic dynamo action. We also show that turbulent heat diffusion is an important part of the energy budget. 

The paper is organized as follows. In \S\ref{section:theory} we review basic theoretical expectations for the interaction between turbulence and magnetic fields. In \S\ref{section:methods} we describe the numerical methods and the setup of the inital conditions. In \S\ref{section:results},
we describe our results, including the level and volume-filling nature of turbulence, evolution of the gas temperature, generation
of vorticity and magnetic fields, and nature of heating mechanisms. Conclusions are summarized in \S\ref{section:conclusions}.\\

\section{Theoretical Expectations: Turbulence and Trapping of g-modes}
\label{section:theory}

It is useful to begin by reviewing some basic theoretical expectations for the behavior of turbulence excited in galactic wakes in clusters. In principle, orbiting galaxies can excite galactic wakes by two means: hydrodynamically (as the ICM collides with the ISM of the galaxy), and gravitationally (similar to dynamical friction for collisionless particles). In practice, we shall conservatively assume that ram pressure stripping is efficient in removing the ISM of galaxies and thus that galaxies only exert gravitational influence. 

{\bf Volume-filling Turbulence} The volume filling factor of galaxy wakes is small (for a simple analytic estimate, see \citet{subramanian06}). This might seem to imply that the impact of turbulence excited by galactic wakes is confined to a small fraction of the cluster. However, orbiting galaxies can also resonantly excite g-modes, which from a formal WKBJ analysis have the dispersion relation: 
\begin{eqnarray}
\omega^{2} = \omega_{\rm BV}^{2} \frac{k_{\perp}^{2}}{k^{2}}. 
\end{eqnarray}
$k^{2}=k_{\perp}^{2} + k_{r}^{2}$. The \brunt frequency for buoyant oscillations is:
\begin{eqnarray}
\left[ (\omega_{\rm BV}^{\rm hydro})^{2}, \omega_{\rm BV}^{\rm MHD})^{2} \right] = \frac{g}{r} \left[ \frac{3}{5} \frac{d\, {\rm ln} S}{d\, {\rm ln} r}, \frac{d\, {\rm ln} T}{d\, {\rm ln} r} \right]
\label{eqn:brunt}
\end{eqnarray}
where $S$ is the fluid entropy $ S \equiv k_{B} T/n^{2/3}$, and $\omega_{\rm BV}^{\rm hydro}$ applies if thermal conduction is negligible, while $\omega_{\rm BV}^{\rm MHD}$ applies if the thermal conduction time is sufficiently short that a displaced blob's temperature is determined by conductive rather than adiabatic cooling \citep{sharma09b}. Note that $\omega_{\rm BV}^{\rm hydro}, \omega_{\rm BV}^{\rm MHD}$ depend on the entropy and temperature gradient respectively; typically, $\omega_{\rm BV}^{\rm MHD}$ is about a factor of 2 smaller than $\omega_{\rm BV}^{\rm hydro}$. 

The above dispersion relation immediately implies that to obtain modes with real $k_{r}$, the driving frequency $\omega < \omega_{\rm BV}$; otherwise the modes have imaginary radial wavenumber and are evanescent. Physically, one can always achieve a low frequency response by making the mode progressively more tangential, but it is impossible to drive the system at frequencies higher than the maximum response frequency of $\omega_{\rm BV}$, corresponding to completely vertical oscillations. This thus implies that waves driven at frequencies $\omega < \omega_{\rm BV}$ can be resonantly excited, and must propagate inward toward the cluster center (as can be seen from their group velocity; \citet{balbus90}), where they will be trapped, reflected and focused inside the resonance radius where $\omega_{\rm BV}=\omega$. A linear analysis by \citet{balbus90} showed that most of the power in $g$-modes is in the longest wavelengths\footnote{Although a WKBJ analysis formally breaks down in this regime, a subsequent numerical study \citep{lufkin95} showed that many of the linear theory results are still valid.}. Note that both $\omega$ (which depends on the orbital frequencies of galaxies) and $\omega_{\rm BV}$ are sensitive to the gravitational potential, which is instrumental in determining if g-modes will be excited. 

{\bf Isotropic Turbulence} Turbulence in the fluid has to compete with buoyancy forces arising from stable stratification. One can show that the ratio of tangential and radial  velocities is given by (e.g., see discussion in \S2 of \citet{ruszkowski10}):
\begin{equation}
\frac{v_{t}}{v_{r}} \sim \left(\frac{\omega_{\rm L}}{\omega_{\rm BV}} \right)^{2} \sim {\rm Fr}^{2}
\end{equation}
where $\omega_{\rm L}=v/L$ is the eddy turnover frequency at a given scale, and Fr is the Froude number, which compares inertial and gravitational forces (${\rm Ri} \sim 1/{\rm Fr}^{2}$ is the Richardson number). If $\omega \ll \omega_{\rm BV}$, then turbulence is fundamentally 2D, and for instance it is difficult to rearrange magnetic fields in the radial direction. However, the level of turbulence required to overcome stable stratification is weak; for typical cluster conditions the critical turbulent velocity is \citep{sharma09b}:
\begin{eqnarray}
\sigma \approx 135 \, {\rm km \, s^{-1}} {g_{-8}}^{1/2} r_{10}^{1/2}\left( \frac{d \, {\rm ln}T/d \, {\rm ln} r}{0.15} \right)^{1/2} \left( \frac{Ri_{c}}{0.25} \right)^{-1/2}\\ \nonumber
\end{eqnarray}
\noindent
where $g_{-8}$ is the gravitational acceleration in units of $10^{-8} \, {\rm cm^{2} s^{-1}}$, $r_{10}$ is a characteristic scale height in units of $10$ kpc, and $Ri_{c}$ is the critical Richardson number; $Ri_{c} \sim 1/4$ is typical for hydrodynamic flow. 

At first blush, the requirement for ${\rm Fr} \gsim 1$ might seem to be at odds with the requirement that $\omega < \omega_{\rm BV}$ for $g$ modes to be excited. However, note that for homogeneous Kolmogorov turbulence, $\omega_{\rm L} \propto L^{-2/3}$; it is therefore conceivable that low frequency g-modes can be excited on large scales, while high-frequency small-scale modes can overcome stabilizing buoyancy forces. Since our background state is not homogeneous, we have to resort to 3D simulations to verify if this expectation is indeed satisfied. This is a major goal of this paper. 
 
{\bf Vorticity and B-field growth} G-modes excite vorticity. An easy way to see this is to examine the vorticity evolution form of the momentum equation for $g$-waves (i.e., assuming $\delta P/P \ll \delta \rho/\rho$) \citep{lufkin95}:
\begin{eqnarray}
\frac{\partial {\bf (\delta \Omega)}}{\partial t} = i \frac{\rho^{\prime}}{\rho} ({\bf k} \times {\bf g}), \\ \nonumber
\label{eqn:vorticity}
\end{eqnarray}
\noindent
where ${\bf \Omega}$ is the vorticity, and to note that ${\bf k}$ is in general non-radial, so that ${\bf k} \times {\bf g}$ is non-zero (indeed, we see in Fig. \ref{fig:frequencies} that since $\omega/\omega_{\rm BV}$ rises toward the center, that g-modes become progressively more tangentially biased there). This implies that vorticity is a good tracer of g-modes, a fact which we shall exploit. It also means that g-modes could conceivably drive an efficient dynamo. There is a well-known analogy between the vorticity equation:
\begin{equation}
\frac{\partial {\bf \Omega}}{\partial t} + \nabla \times ({\bf \Omega} \times {\bf u})= - \nabla \times (\nu \nabla \times {\bf \Omega})
\end{equation}
where $\nu$ is the viscosity, and the relation for the magnetic field in the flux-freezing limit,
\begin{equation}
\frac{\partial {\bf B}}{\partial t} + \nabla \times ({\bf B} \times {\bf u})= - \nabla \times (\eta \nabla \times {\bf B})
\end{equation}
where $\eta$ is the electricity resistivity. This, together with the fact that the divergence of ${\bf B}$ and ${\bf \Omega}$ both vanish, leads to the expectation that their growth might be related\footnote{Note, however, that this analogy is imperfect, since ${\bf \Omega} = \nabla \times {\bf u}$, which leads to a nonlinear coupling in the equations, whereas no such relation exists between ${\bf B}$ and ${\bf u}$.}. There have been a number of studies pointing out that turbulent motions could give rise to magnetic fields in clusters (e.g., \citet{ruzmaikin89,subramanian06,ryu08,cho09}). This subject is rich and beyond the scope of this paper; we shall merely compare the growth of vorticity and magnetic fields in our simulations, to see how well they track one another. A reasonable expectation is that the magnetic fields achieve equipartition with turbulence (e.g., \citet{schekochihin07}, and references therein): 
\begin{equation}
B_{\rm eq} \approx 7 \mu{\rm G} \left( \frac{n_{e}}{0.02 \, {\rm cm^{-1}}} \right)^{1/2} \left(\frac{v_{\rm turb}}{100 \, {\rm km \, s^{-1}}} \right).
\label{eqn:B_field_equi}
\end{equation}
where $v_{\rm turb}$ is the rms turbulent velocity on large scales. The above estimate is consistent with observed $\sim \mu$G fields \citep{carilli}, though there are considerable uncertainties\footnote{In general, estimates based on rotation measure (RM) lead to stronger magnetic fields, while
those based on synchrotron and inverse Compton (IC) analysis give weaker fields. However, RM methods may overestimate fields
if single-scale magnetic field correlation length is used \citep{newman} or when the small-scale fluctuations in density and magnetic field
are correlated in a turbulent medium \citep{beck}. Moreover, these estimates depend on whether 
radio sources used to probe the field strength are embeded in the ICM, with smaller values inferred when background sources rather than embedded
ones are used \citep{carilli}.}. The fact that trapping of $g$-modes can give rise to volume-filling turbulence would then be instrumental in allowing volume-filling magnetic fields. 

{\bf Magnetic tension} Magnetic tension can inhibit the HBI \citep{quataert08}.  
For perturbation scales comparable to the radius $r$ (i.e., $\lambda =2r$) we obtain a critical value: 
\begin{eqnarray}
B_{\rm crit} \gsim & & 10  \mu{\rm G},
\left( \frac{g}{10^{-7} \, {\rm cm \, s^{-2}}} \right)^{1/2} \left( \frac{n_{e}}{0.02 \, {\rm cm^{-3}}} \right)^{1/2} \label{eqn:B_field_hbi}  \\ \nonumber
   & & \left( \frac{r}{30 \, {\rm kpc}} \right)^{1/2} \left( \frac{d{\rm ln \, T}/d{\rm ln r}}{0.3} \right)^{1/2} 
\end{eqnarray} 
for suppression, where the fiducial values are measured from our simulated cluster at a radius of $r=30$kpc. Due to the similarity between the field values in equation (\ref{eqn:B_field_equi}) and (\ref{eqn:B_field_hbi}), it has been suggested that magnetic fields amplified by turbulence can prevent the onset of the HBI (e.g., see discussion in \citet{kunz10}. Note that their version of equation
(\ref{eqn:B_field_hbi}) yields somewhat lower B-fields than ours, for identical parameters. In any case, equation (\ref{eqn:B_field_hbi}) is only an approximation as the derivation assumes the  
WKB approximation, while the non-linear saturation of the HBI occurs on global scales). In this paper, we will deliberately ignore this possibility.   
Observationally, the strength of the magnetic field in the ICM is $\sim\mu$G and has a large scatter of about an order of magnitude 
within the ICM and between clusters \citep{carilli}; moreover, there are considerable observational uncertainties in these values, as mentioned above. Numerical simulations show that the HBI still develops for $\sim\mu$G fields
(I. Parrish, priv. comm.), although it can be delayed for increased field strengths. Given the large uncertainty in whether observed field strengths are capable of stabilizing the HBI, past studies of HBI (e.g., \citet{parrish09, bogdanovic09, ruszkowski10, parrish10}) 
focused on the regime where the magnetic tension
is unimportant. We also adopt the same approach here, and study if volume-filling turbulence {\it alone} can stabilize the HBI. More specifically, we consider plasma $\beta\gg 1$ and note that, as long as the 
field is not dynamically important, its exact value does not play a role. In this case, the magnetic field strength scales out of the problem and only serves as a medium to redirect the heat flow via anisotropic thermal conduction. We can therefore study the effects of turbulence alone without the possibly confounding effects of magnetic tension.  

{\bf Turbulent heating and heat diffusion} Turbulence impacts the thermodynamics of the fluid through its effect on thermal conduction, both randomizing and amplifying the magnetic field. Both of these suppress the HBI, and allow thermal conduction at $\sim 1/3$ the Spitzer value. However, turbulence can also directly affect the thermal state of the plasma through dissipation of turbulent motions (direct heating), or allowing heat transport via turbulent diffusion (\citet{kim03,dennis05}, and references therein). The heating rate from dissipation of kinetic and magnetic energy is: 
\begin{equation}
\Gamma_{\rm diss} = \frac{c_{\rm diss} \rho u^{3}}{l}
\end{equation}
where $c_{\rm diss}$ is a dimensionless constant of order unity and $l$, the dominant velocity length-scale, is unknown but almost certainly a function of radius; a reasonable ansatz might be $l\approx \alpha r + l_{0}$ \citep{dennis05}, where $\alpha$ is some adjustable constant of order unity, and $l_{0}$ is some minimal lengthscale. On the other hand, the heating rate from turbulent heat diffusion is: 
\begin{equation}
\Gamma_{\rm diff} = \nabla \cdot (\kappa_{\rm turb} \rho T \nabla s)
\label{eqn:heat_diffusion}
\end{equation}
where $s=C_{\rm V} {\rm ln}(p/\rho^{\gamma})$ is the specific entropy, and the turbulent diffusivity is: 
\begin{equation}
\kappa_{\rm turb} \approx u l \ {\rm min} \,  (1, \left(\frac{\omega}{\omega_{\rm BV}}\right)^{2})
\end{equation}
where the second factor of $(\omega/\omega_{\rm BV})^{2}$ takes into account the damping of radial heat transport by buoyancy forces \citep{dennis05}. The fact that $\kappa_{\rm turb} \sim u l$ is of order the hydrodynamic value even for a magnetized plasma was found in MHD simulations by \citet{cho03}. Nonetheless, equation (\ref{eqn:heat_diffusion}) should be understood to be only approximate, since it assumes that fluid elements are transported adiabatically, which need not be the case when anisotropic conduction is operating. In reality, both the thermal conduction diffusion coefficient $\kappa_{\rm Spitzer} = v_{e} l_{\rm mfp} \sim 10^{30} {\rm cm^{2} \, s^{-1}} \, (n/10^{-2} \, {\rm cm^{-3}})^{-1} (T/2 \, {\rm keV})^{5/2}$ and the turbulent heat diffusion coefficient $\kappa_{\rm turb} \sim 10^{30} \, {\rm cm^{2} \, s^{-1}} \, (u/200\, {\rm km \, s^{-1}})(l/20 \, {\rm kpc})$ can be comparable, and either could dominate in a specific situation. 

Thermal conduction may indirectly assist with turbulent heat diffusion, as it reduces the impact of buoyancy forces (and thus reduces $\omega_{\rm BV}$). Indeed, simulations by \citet{sharma09a} show that metal mixing in a stratified plasma is much more efficient once conduction is at play, allowing much broader metallicity profiles, for this very reason. Naively, if we think of gas entropy as a scalar to be advected by turbulent motions, similar conclusions should hold, although of course the interaction between heat transport and dynamics requires detailed simulations. We shall investigate the relative role of all these heating processes in our simulations. 

\section{Methods}
\label{section:methods}

\subsection{Initial conditions for the gas}
\label{section:initial_conditions}

The details of the numerical setup are described in Ruszkowski \& Oh (2010; hereafter RO10). Here we summarize key differences.
The cluster parameters used here are similar to those corresponding to cool-core cluster A2199. In addition to the NFW potential 
of the cluster halo, we also include the contribution from the central brightest cluster galaxy (BCG), which was not included in RO10. The gravitational potential is described by the sum
of the term due to an NFW profile with a softened core\\
\begin{eqnarray}
\Phi & = & - 2GM_{0}  \frac{r_{c}}{(r_{s}-r_{c})^{2}}\left [ \ln\frac{1+r/r_{c}}{1+r/r_{s}}   + \frac{\ln(1+r/r_{c})}{r/r_{c}}\right] \nonumber \\
     &   & - 2GM_{0}\frac{r_{s}(r_{s}-2r_{c}) }{r_{c}(r_{s}-r_{c})^{2}}\frac{\ln (1+r/r_{s})}{r/r_{c}}  ,\\ \nonumber 
\label{eqn:NFW_potential}
\end{eqnarray}
\noindent
where $r_{c}$ is the smoothing core radius ($r_{c}= 20$ kpc), $r_{s}=390$ kpc  
is the usual NFW scale radius, and the BCG contribution which has a King profile: 
\begin{eqnarray}
\Phi_{\rm bcg} & = & -9\sigma_{\rm bcg}^{2}\left[\frac{\ln \left(x + \sqrt{1+x^{2}}\right )}{x} \right], \\ \nonumber 
\end{eqnarray}
\noindent
where $x=r/r_{\rm bcg}$,
$r_{\rm bcg}=3$ kpc is the core radius for the BCG and $\sigma_{\rm bcg}=200$ km s$^{-1}$ is 
its line-of-sight velocity dispersion.  
The parameter $M_{o}=3.8\times 10^{14}M_{\odot}$ in equation (\ref{eqn:NFW_potential})
determines the cluster mass and is of the order of the total cluster mass, $M_{\rm 200}=6.6 \times 10^{14} \, M_{\odot}$.  
We then solve the equation of hydrostatic equilibrium assuming the entropy distribution as parametrized by \citet{cavagnolo09}; see equations (16) \& (17) of RO10. Note that we do not include the gravitational contribution from other galaxies (\S \ref{section:galaxies}) in our initial conditions, so the system is not initially in full hydrostatic equilibrium. However, after an initial transient, it rapidly relaxes to a new equilibrium configuration. 

The addition of the BCG has two effects. Firstly, due to the increased gravitational acceleration, it results in higher gas densities compared to the models we considered in Ruszkowski \& Oh (2010). This allows for a more conservative analysis of the effect of cooling. In fact, the central density here is a factor of $\sim 3.5$ times higher, which, combined with a slightly lower assumed central temperature, results in a central cooling time which is nearly 5 times shorter. The higher adopted central density in this paper is in line with that observed in A2199 \citep{johnstone02}. Given this more stringent setup, some of the stable models in Ruszkowski \& Oh (2010) would actually undergo a cooling catastrophe. 
Secondly, the change in the gravitational potential has consequences for the \brunt frequency and the trapping of g-modes, as we discuss below. 

\begin{figure}
\includegraphics[width=0.45\textwidth, height = 0.45\textwidth]{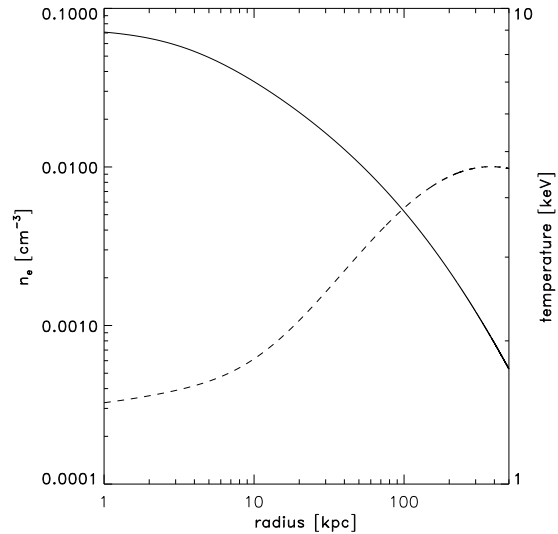}
\caption{Initial electron number density (solid line) and temperature (dashed line) in the ICM of our simulated cluster.}
\label{plot1}
\end{figure}

\begin{figure}
\includegraphics[width=0.45\textwidth, height = 0.45\textwidth]{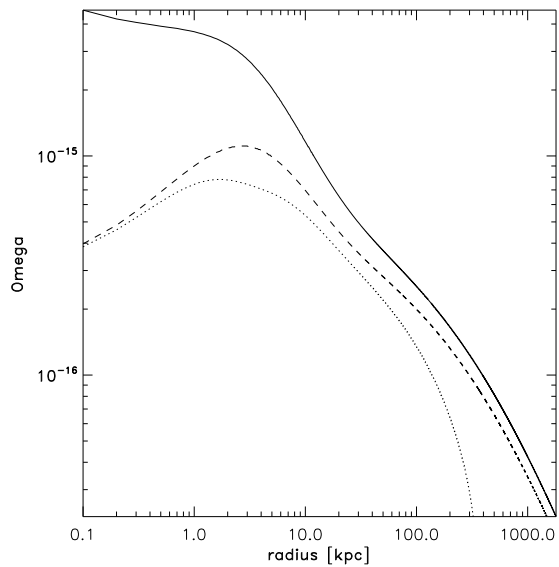}
\caption{The frequency of circular orbits $\omega_{\rm orb}$ (solid line), the \brunt frequency $\omega_{\rm BV}^{\rm hydro}$
(dashed line) for a hydrodynamic fluid and $\omega_{\rm BV}^{\rm MHD}$ (dotted line) 
for a magnetized conducting fluid (all in [Hz]). The frequencies correspond to the initial density and temperature profile shown in Figure 1.}
\label{fig:frequencies}
\end{figure}

The initial distribution of density and temperature is shown in Figure \ref{plot1}.
The frequency of circular orbits $\omega_{\rm orb}$ and the \brunt frequencies $\omega_{\rm BV}^{\rm hydro}, \omega_{\rm BV}^{\rm MHD}$ for a hydrodynamic and conducting fluid with this initial density and temperature profile are shown in Figure \ref{fig:frequencies}. 
For a mode with a given value of $\omega$, $g$-modes can be resonantly excited if $\omega < \omega_{\rm BV}$. This therefore defines an outer trapping radius for such a mode. Note that both $\omega_{\rm orb}$ and $\omega_{\rm BV}$ are both strong functions of the gravitational potential. We have directly verified the resonance condition by running simulations both with and without the central cD galaxy; in the latter case, orbiting galaxies fail to excite volume-filling turbulence, which is to be expected since $\omega_{\rm BV}$ falls inward in this case, and the resonance condition is never satisfied (see also \citet{lufkin95,kim07}). Note that fine-turning of the resonance condition is not necessary: the resonance is not very sharp \citep{balbus90}, and in practice galaxies with non-circular orbits excite modes with a variety of harmonics, some of which can potentially fall below $\omega_{\rm BV}$. 

The magnetic field setup was identical to that in Ruszkowski \& Oh (2010): we generate statistically isotropic random-phase complex fields in Fourier space, with 3D Fourier amplitudes given by: 
\begin{eqnarray}
B_{k} \propto k^{-11/6}\exp\left[-\left(\frac{k}{k_{o}}\right)^{4}\right],
\end{eqnarray}
as appropriate for Kolmogorov turbulence, where $k_{o}=2\pi/\lambda_{o}$ and $\lambda_{o}\sim 43h^{-1}$ kpc is a smoothing wavelength. We then apply a divergence cleaning operator in ${\bf k}-$space, and then inverse Fourier transform the field back to real space. 

\subsection{Initial conditions for the galaxies}
\label{section:galaxies}

The simulations must be initialized with a galaxy population, which has the appropriate spatial distribution, masses and velocities. Rather than relying upon cosmological simulations, we use an empirically grounded approach, which also has the advantage of speed and flexibility. How are the galaxies spatially distributed? From a sample of K-band selected galaxies within 93 clusters and groups, \citet{lin04} find that the galaxy number density profile in clusters is well described by the NFW profile \citep{navarro97} with a concentration parameter $c \sim 3$, with no evidence for cluster mass dependence of the concentration. The theoretical justification for galaxies tracing the NFW profile is somewhat equivocal. If one attempts to use cosmological simulations to set up initial conditions, the radial distribution of subhalos in simulations is well-known to be less concentrated than the dark-matter, or 'anti-biased' (\citet{nagai05}, and references therein). This is due to tidal stripping of sub-halos in the central regions. On the other hand, simulations that include galaxy formation allow subhalos to be selected by stellar mass. This generally shows closer agreement with observed profiles, as the stellar mass (which is tightly bound) remains conserved while the dark matter is stripped from outer regions \citep{nagai05}. Other investigators find that the fraction of such stellar-dominated halos is small, but caution that numerical resolution effects may preclude robust conclusions at this point \citep{dolag09}. Overall, we therefore simply employ the observational result that galaxies trace the NFW profile. 

As for the galaxy masses, instead of using a Schechter function, we simply assume (as did, for instance, \citet{subramanian06,kim07}) that all galaxies have the same mass. This is for two reasons. Firstly, this allows us to rapidly explore the effect of varying galaxy masses (due, for instance, to different efficiencies of tidal stripping). The assumption of a characteristic mass is reasonable: since dynamical friction scales as $M_{\rm gal}^{2}$, most turbulent motions are induced by galaxies of mass $\sim M_{*}$, where most of the mass resides, rather than the more abundant lower mass galaxies. Indeed, we shall find that the induced gas motions are mostly sensitive to the mass of galaxies, and less sensitive to their number (\S\ref{section:velocities}). Previous hydrodynamic simulations found unchanged results with galaxies drawn from a Schechter distribution, if the characteristic break mass $M_{*} \sim M_{\rm gal}$ \citep{kim07}. Secondly, it allows us to directly calibrate against lensing estimates for subhalo mass fraction. Unlike K-band surveys, lensing is directly sensitive to total mass, but is generally only sensitive to subhalos with $M \gsim 10^{11} \, M_{\odot}$. \citet{natarajan09} find from the massive lensing cluster Cl 0024+16 that $\sim 30\%$ of the cluster mass can be attributed to substructure with $M \gsim 10^{11} \, M_{\odot}$, with typical masses $\sim 10^{12} \, M_{\odot}$ (with a weak radial trend such that galaxies in the outer regions are more massive; see their Fig 6). Their results, including the mass function as a function of radius, is broadly consistent with the results of the Millenium simulation run, except that the typical masses of galaxies is lower in simulations by a factor of $\sim 3$. This is subject to the uncertainties of extra binding due to a compact stellar halo mentioned above; note that masses of $\sim 10^{12} \, M_{\odot}$ is also consistent with other observations from lensing \citep{shin08} and galaxy wakes \citep{sakelliou05}. Below we explore a grid of models with varying galaxy mass, but never allowing the total substructure mass fraction to rise above $\sim 25\%$. For simplicity in the code, the galaxies are modeled as point masses. Since we are primarily concerned with the excitation of g-modes on scales much larger than $\sim$kpc galactic scales, we do not expect this simplification to significantly impact our results. 

Given these assumptions, the most rigorous way to initialize galaxy velocities is to directly construct the
distribution function from the density profile, using Eddington's
formula \citep{binney08,kazantzidis04}. However, velocity anisotropy
is only easily incorporated in such models if it has certain
parametric forms, as for instance in Osipkov-Merritt models. Instead,
we construct a self-consistent velocity model via the local Maxwellian
approximation: approximating the velocity dispersion tensor by a
multi-variate Gaussian at each point, with dispersions given by the
solution of the Jeans equation \citep{hernquist93}. This has the
virtue of simplicity and flexibility. Note that such models may not be
in strict equilibrium, and can demonstrate evolution
\citep{kazantzidis04}. However, \citet{springel05a} find the actual
amount of relaxation to be small; furthermore, \citet{faltenbacher05},
who directly simulate the motion of galaxies in clusters, find their
velocity distribution is indeed closely Maxwellian, with good
agreement between simulation results and equilibrium Jeans equation
solutions.
We therefore solve the Jeans equation assuming no rotational support or bulk streaming ($\bar{v_{r}}=\bar{v}_{\phi}=\bar{v}_{\theta}=0$):

\begin{eqnarray}
\frac{1}{n_{\rm gal}} \frac{\rm d}{{\rm d}r} \left( n_{\rm gal} \sigma_{r}^{2} \right) + 2 \beta_{v} \frac{\sigma_{r}^{2}}{r} = -  \frac{\rm d \phi}{{\rm d}r}\\\nonumber
\label{eqn:jeans}
\end{eqnarray}

\noindent
where $\beta_{v}$ is the velocity anisotropy 
parameter\footnote{From cosmological simulations, \citet{benson05} finds that radial 
and tangential velocities can be correlated (at least at the time of merger), a detail we ignore.}:

\begin{eqnarray}
\beta_{v}(r)=1 - \frac{\sigma_{t}^{2}(r)}{2 \sigma_{r}^{2}(r)},\\ \nonumber
\end{eqnarray}

\noindent
$n_{\rm gal}$ is the galaxy number density and $\phi$ is the combined cluster + cD galaxy
gravitational potential. Note that we have not built self-consistent models and ignore the contribution of galaxies to the gravitational potential; for a large sub-halo mass fraction, the system is not in full equilibrium. In practice, this is a small effect, and the galaxy distribution does not evolve significantly over the course of our simulation.\\
\indent
What are appropriate assumptions for
$\beta_{v}(r)$? It may be estimated from observations via Jeans equation
modelling, given knowledge of galaxies positions, the cluster
potential, and line of sight velocities. A detailed study of 10
clusters using a spectroscopic sample of galaxies from SDSS and 2dF
found galaxy orbits to be isotropic within the errors for most
clusters \citep{hwang08}. An earlier paper, using ENACS data, found
that the brightest ellipticals do not yield an equilibrium solution,
while other ellipticals, SOs and early spirals have isotropic orbits, and 
late spirals prefer radial to isotropic orbits
\citep{biviano04}. Overall, we assume isotropy $\beta_{v}(r)=0$, and
regard this as our default model. In passing, we note that one could
easily incorporate the effects of the velocity anisotropy by, for example,
considering fits to to measurements in simulations (\citet{hoeft04}).
Given that the evidence for orbital anisotropy in observations is marginal to date, we defer the study of the effect of such orbital distributions to future work. 
We also note that preferentially radial orbits would enhance the restoration of conduction even further and strengthen our conclusions. 

We solve equation (\ref{eqn:jeans}) as an initial value problem, where $\sigma_{r}^{2}(r_{\rm 200}) \approx GM_{200}/3r_{200}$. Having solved for $\sigma_{r}(r)$ and $\sigma_{t}(r)$, we randomly sample from the
multi-variate Gaussian at each position to create a realization of the
velocity field. The above procedure allows us to initialize the
simulations with a realization of galaxies with both masses and six-dimensional
phase space coordinates (position and velocity).\\

\begin{figure*}
\includegraphics[width=0.45\textwidth]{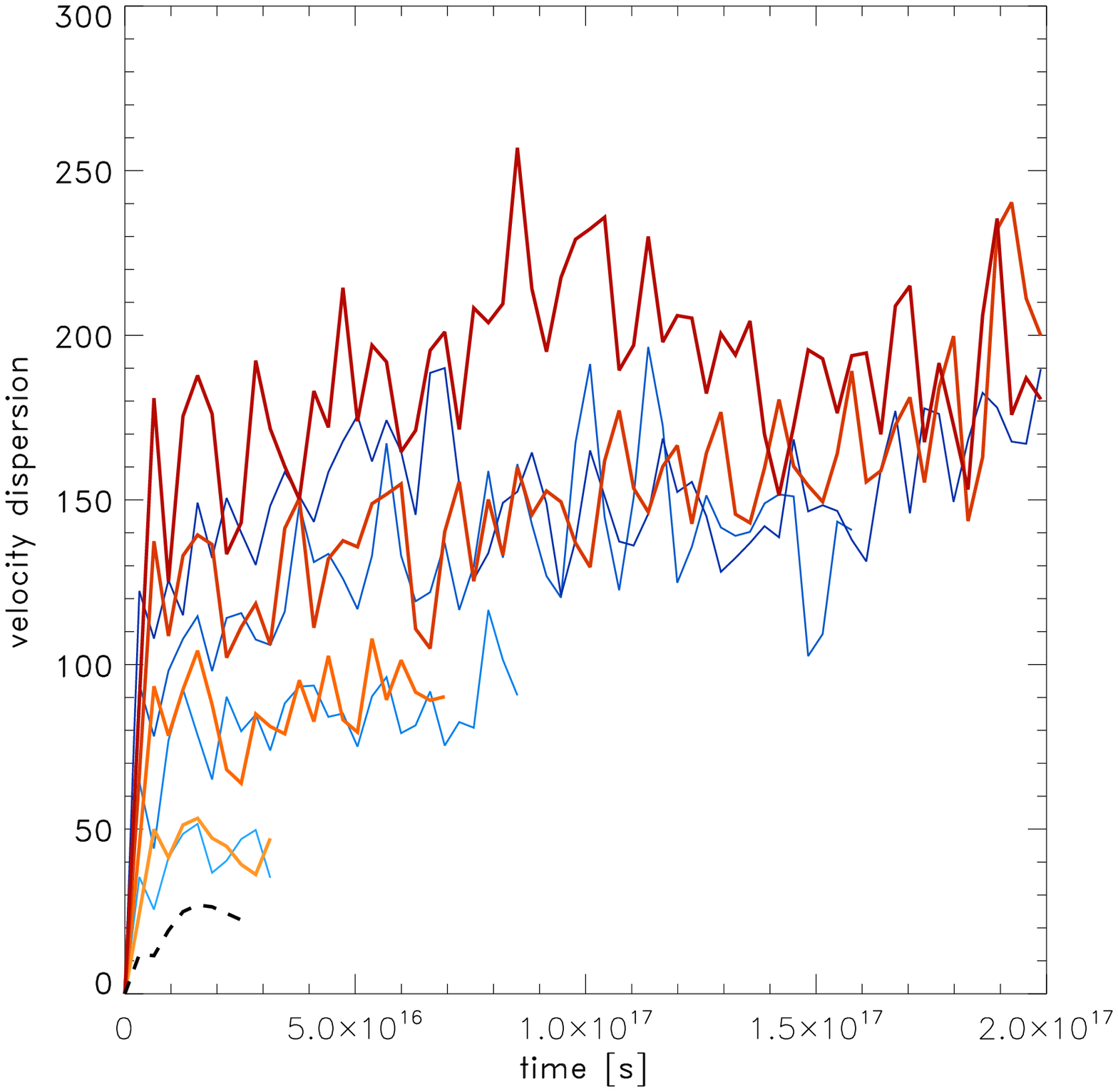}
\includegraphics[width=0.45\textwidth]{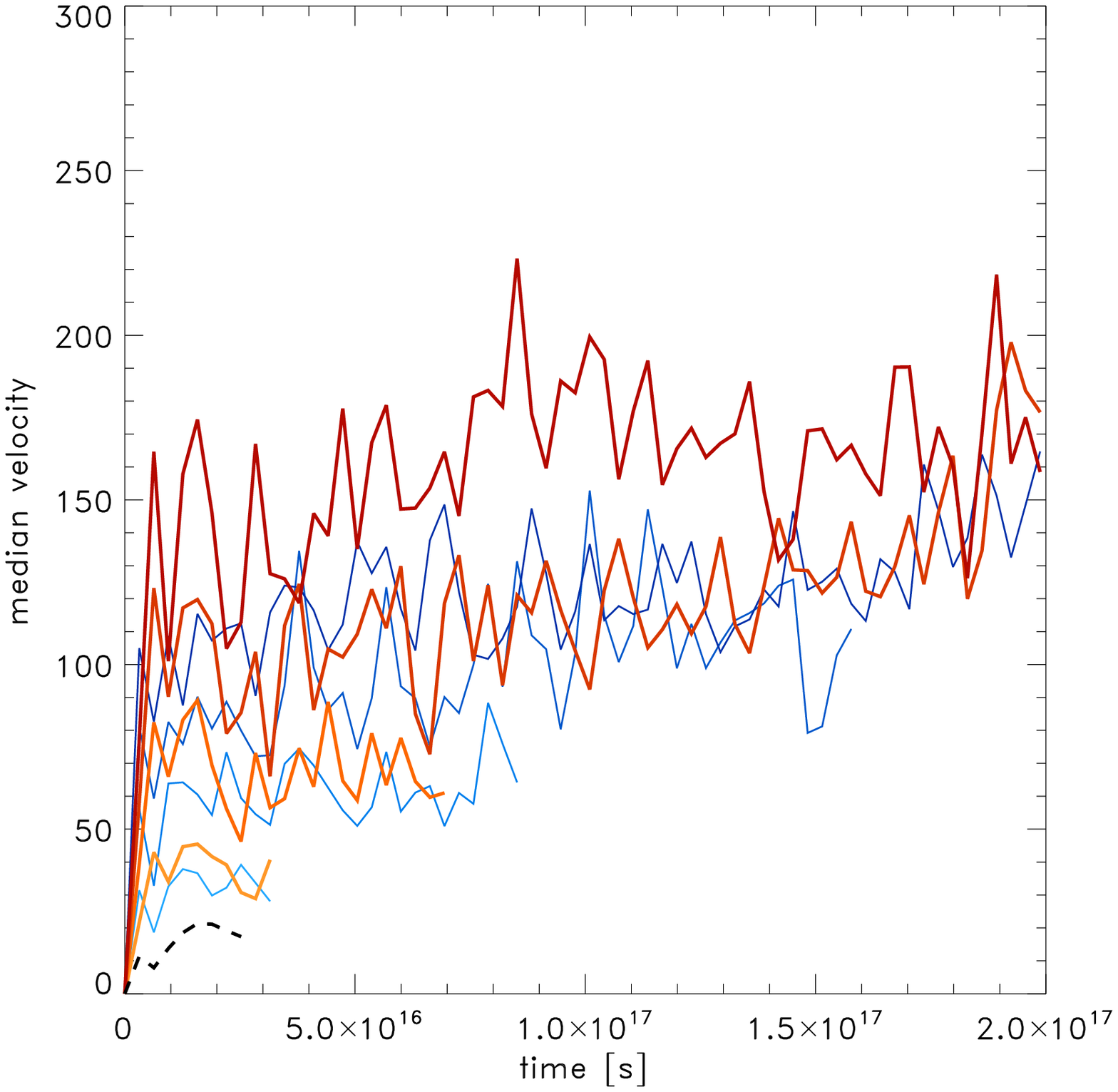}
\caption{The evolution of the velocity dispersion (left panel) and median velocity within the central 100 kpc (all in [km s$^{-1}$]). 
Blue (red) lines are for 100 (200) galaxies respectively, for equally spaced masses ranging from $3\times 10^{11}$M$_{\odot}$ 
to $1.2\times 10^{12}$M$_{\odot}$. The galaxy mass increases gradually from the lightest  to the darkest color. 
The black dashed line is for the pure HBI case, where there is no stirring. The run is halted at early times if a cooling catastrophe occurs.}
\label{fig:velocities}
\end{figure*}

\subsection{Simulation}

The simulations were performed using the {\it FLASH} code (version 3.2). {\it FLASH} is a modular, parallel adaptive mesh refinement magnetohydrodynamic 
code. Magnetic field evolution was solved by means of a directionally
unsplit staggered mesh algorithm (USM; \citet{lee09}). 
The USM module is based on a finite-volume, high-order Godunov scheme combined with constrained
transport method (CT). This approach guarantees divergence-free magnetic field distribution. We implemented the anisotropic conduction unit
following the approach of \citet{sharma07}. More specifically, we applied monotonized central (MC) limiter to the conductive fluxes.
This method ensures that anisotropic conduction does not lead to negative temperatures in the presence of steep temperature 
gradients. 
The three-dimensional computational domain was approximately 1 Mpc on each side, 
enclosing a large fraction of the cluster. 
The central regions of the cluster had an enhanced refinement level. The maximum spatial resolution for 6 levels of refinement was 
$\sim 2.7h^{-1}$ kpc.
The simulations were performed on a 384-processor cluster located at the Michigan Academic Computing
Center at the University of Michigan in Ann Arbor and on the {\it Columbia} supercomputer at NASA Ames.\\

\section{Results}
\label{section:results}

We performed a total of 16 runs including radiative cooling, anisotropic thermal conduction and self-gravitating particles 
to emulate the gas ``stirring'' by galaxies. We considered a uniform grid of parameters: 50, 100, 150, and 200  galaxies characterized by masses of 
(0.3, 0.6, 0.9, 1.2) $\times 10^{12}$ M$_{\odot}$. With our cluster mass of $6.6 \times 10^{14} \, {\rm M_{\odot}}$, these parameters corresponds to a mass fraction in galaxies ranging from $f_{\rm gal}=2.2\%$ to a maximum of $f_{\rm gal}=27\%$. For instance, for galaxies with mass $6 \times 10^{11} \, {\rm M_{\odot}}$, our grid corresponds to $f_{\rm gal}=(4.3, 8.3, 12, 15) \%$. We also performed two control runs: one without the galaxies (and hence without stirring) to isolate the effect of 
heat buoyancy instability, and one without conduction to isolate the effect of dynamical friction heating by galaxies.  
 
\subsection{Gas velocities and volume filling of turbulence}
\label{section:velocities} 

Figure \ref{fig:velocities} (left panel) shows the evolution of the velocity dispersion measured within 100 kpc from the cluster center. Thin blue (red) lines are for 
100 (200) galaxies respectively 
and for equally spaced masses ranging from $3\times 10^{11}$ to $1.2\times 10^{12}$. The mass increases gradually from the lightest 
to the darkest color. The black dashed line is for the pure HBI case. The HBI case and lighter-colored curves are evolved for shorter times. These runs suffer from overcooling and the central temperatures
reaches the low temperature threshold at which point the simulation is stopped. The right panel in Figure  \ref{fig:velocities} shows the median velocity within 100 kpc; the
color coding corresponds to that in the left panel. It is clear from these figures that there is a clear trend for the velocity dispersion or the median velocity
to increase with the typical galaxy mass. A similar, albeit weaker, trend is seen for the galaxy number. This is consistent with the findings of 
Kim (2007) who found in pure hydrodynamic simulations, that the gas velocity dispersion $\sigma$ scales as 
$\sigma\propto N_{\rm gal}^{1/2}M_{\rm gal}$, where $N_{\rm gal}$ and $M_{\rm gal}$ are the number and mass of galaxies, respectively. Note that a scaling ${\rm E}_{\rm k} \propto \sigma^{2} \propto N_{\rm gal} M_{\rm gal}^{2}$ is consistent with dynamical friction in the linear regime, since $\dot{\rm E}_{\rm k} \propto M_{\rm gal}^{2}$ for dynamical friction. 

As $N_{\rm gal}$ and $M_{\rm gal}$ increase, the cooling catastrophe is delayed, and is completely staved off at the upper envelope of these parameters. In this respect our MHD simulations differ markedly from those of \citet{kim07}, who found that a cooling catastrophe was inevitable in purely hydrodynamic simulations, for all portions of parameter space.  We explore these differences further in \S\ref{section:heating}.  
Note that in our case the velocity dispersion seems to increase more slowly than $\sigma \propto N^{1/2}$. Besides the inclusion of MHD in our simulations, differing results could be due to a variety of factors, including the different assumed distribution of galaxies. 
Note that the stated number of galaxies are distributed over the entire cluster; the number of galaxies in the inner regions which actually result in trapped g-modes is actually considerably smaller, and subject to Poisson fluctuations. Also, the introduction of more galaxies and/or increasing their mass does not cause the velocity dispersion to increase without limit; instead, the growth in velocity dispersion appears to saturate. \citet{kim07} also observed this in his hydrodynamic simulations, and attributed it to loss of resonant excitation once density fluctuations become large and the background is nonlinear. We see the same saturation on the same $\sim 10^{8}$yr timescale, in simulations with driven volume-filling turbulence, where resonant excitation of modes is not an issue (see \S 3.4 of RO10).  
The asymptotic velocities of $\sim 100-200 \, {\rm km \, s^{-1}}$, while generally insufficient for turbulent heating to be important, is enough to restore thermal conduction and enable turbulent heat diffusion. \\
\indent
The comparison between
the velocity dispersion and median velocity reveals that that both of these quantities are comparable, as might be expected for volume filling turbulence. 

\subsection{Bias in magnetic field orientation}
\label{section:anisotropy}
\begin{figure}
\includegraphics[width=0.45\textwidth, height = 0.45\textwidth]{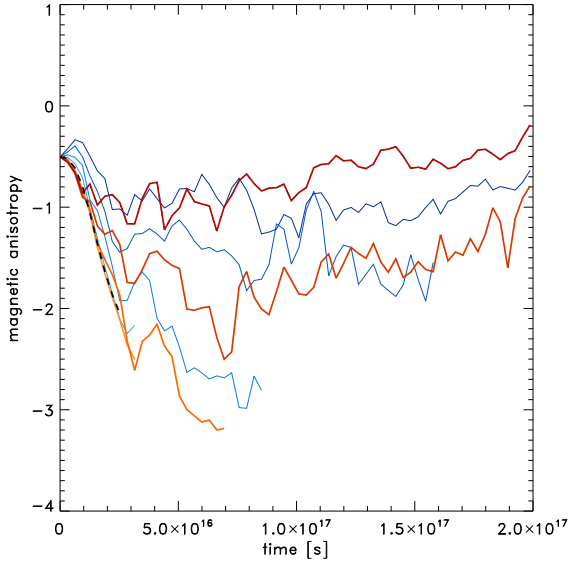}
\caption{The evolution of the anisotropy $\beta$ in the orientation of magnetic fields. Vanishing 
$\beta$ corresponds to isotropic fields. 
The more negative $\beta$ becomes, the more tangential the fields are.
Thin blue lines are for 100 galaxies and for equally spaced masses ranging from $3\times 10^{11}$M$_{\odot}$ 
to $1.2\times 10^{12}$M$_{\odot}$. The galaxy mass increases gradually from the lightest  to the darkest color. 
Thick red curves are the corresponding lines for 200 galaxies. The black dashed line is for the pure HBI case. Runs are terminated when a cooling catastrophe sets in.}
\label{fig:beta_mag}
\end{figure}

The evolution of the anisotropy $\beta$ in the orientation of magnetic fields is shown in Fig. \ref{fig:beta_mag}.
The definition of this parameter is similar to that for galaxy velocity $\beta_{v}$ defined in Eq. (16). The only difference is that 
the velocity dispersions are replaced by magnetic field dispersions. Thus, $\beta=0$ corresponds to isotropic
magnetic fields, whilst $\beta \rightarrow (-\infty,1)$ corresponds to progressively more tangential (radial) fields respectively.
Thin blue lines are for 100 galaxies and for equally spaced masses ranging from $3\times 10^{11}$M$_{\odot}$ 
to $1.2\times 10^{12}$M$_{\odot}$. The galaxy mass increases gradually from the lightest  to the darkest color. 
Thick red curves are the corresponding lines for 200 galaxies. The black dashed line is for the pure HBI case.
As expected, when stirring is weak, the HBI prevails, leading to a systematic tangential bias in the orientation of magnetic fields. This insulates the core against thermal conduction, leading to a cooling catastrophe. The fields become more tangential with time and the cluster eventually suffers from overcooling. 
On the other hand, for increasingly vigorous stirring (i.e., increasing the individual masses or number of galaxies), the field becomes increasingly isotropic, and a cooling catastrophe is averted. 

These results are consistent with the driven turbulence simulations in RO10\footnote{Although note that all but one of the simulations in RO10 were adiabatic simulations; by contrast, all the simulations presented here simultaneously include radiative cooling.}, and can be broadly understood in terms of the simple Froude/Richardson number criterion outlined in \S\ref{section:theory}. The main difference in the more realistic scenario we present here is that the discrete nature of the stirrers and resonant excitation process introduces greater stochasticity and time-dependent fluctuations in the velocity field and magnetic anisotropy (e.g., compare the smooth curves in Fig 3 \& 4 of RO10 with their noisier equivalents Fig \ref{fig:velocities} \& \ref{fig:beta_mag} of this paper). But the main physical conclusions are unchanged. It is also interesting to note that while the velocity dispersion is only weakly dependent on the number of galaxies (depending more sensitively on galaxy masses), the magnetic anisotropy shows somewhat greater sensitivity. In particular, the magnetic field anisotropy cannot simply be predicted from the instantaneous velocity dispersion, as in a naive application of a Froude/Richardson criterion. We saw similar behavior in RO10, where runs with similar asymptotic velocity dispersions had similar velocity anisotropies, but markedly different magnetic anisotropies. The advected magnetic field is sensitive to the integrated past displacement history of a fluid element, and not merely the instantaneous velocity field.    

\subsection{Evolution of Gas Temperature and Entropy}
\label{section:temperature} 
\begin{figure*}
\includegraphics[width=0.33\textwidth]{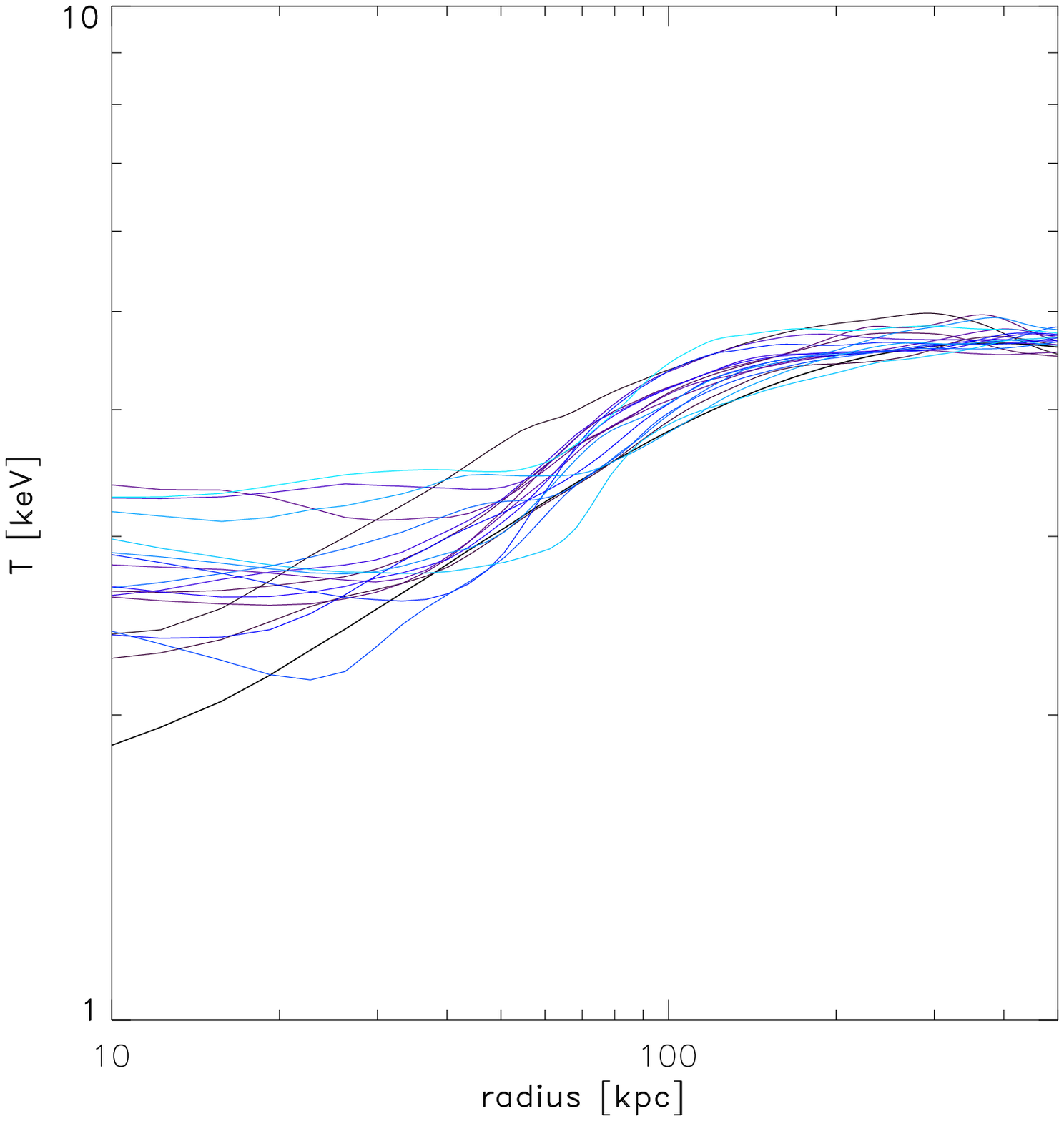}
\includegraphics[width=0.33\textwidth]{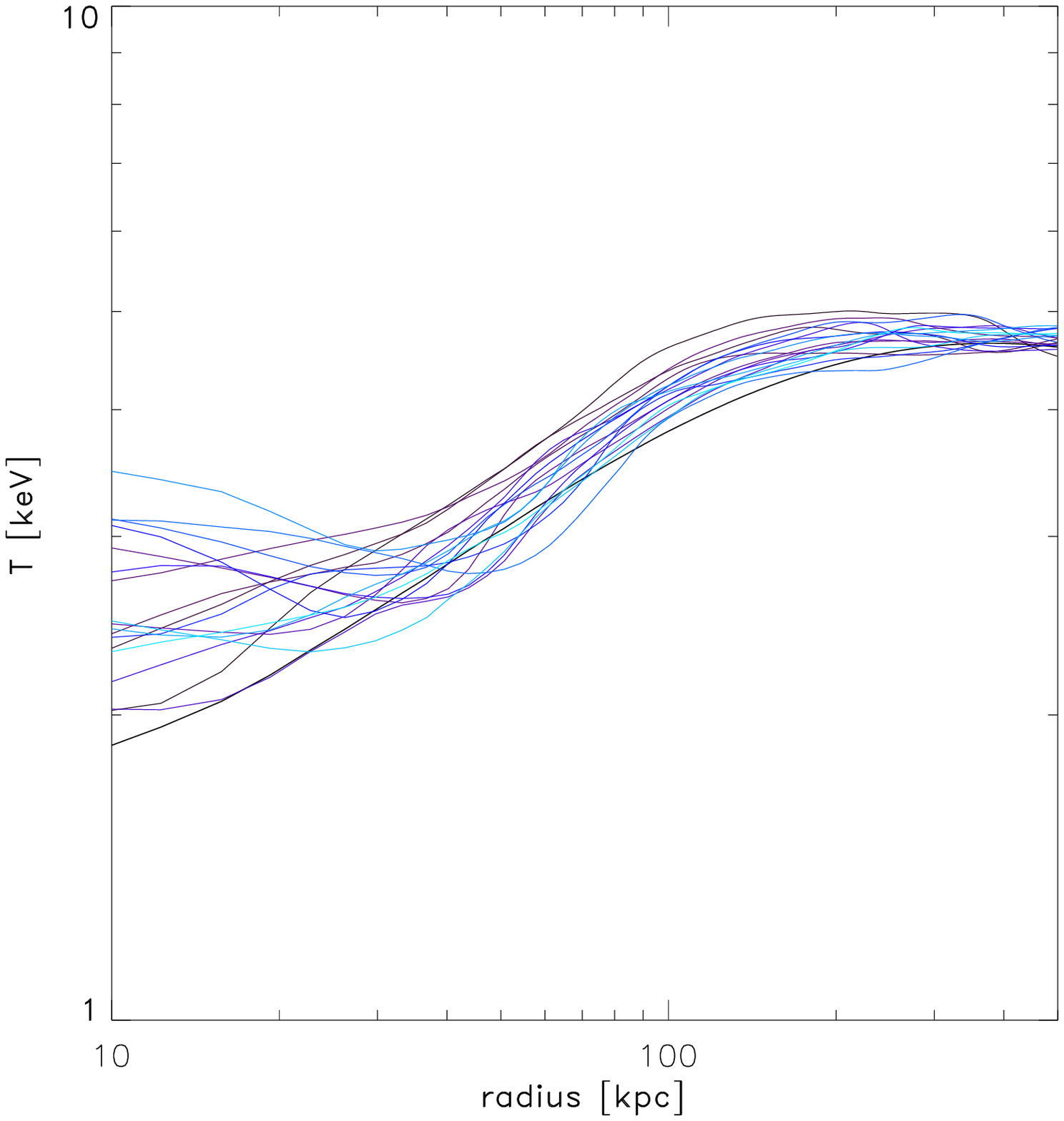}
\includegraphics[width=0.33\textwidth]{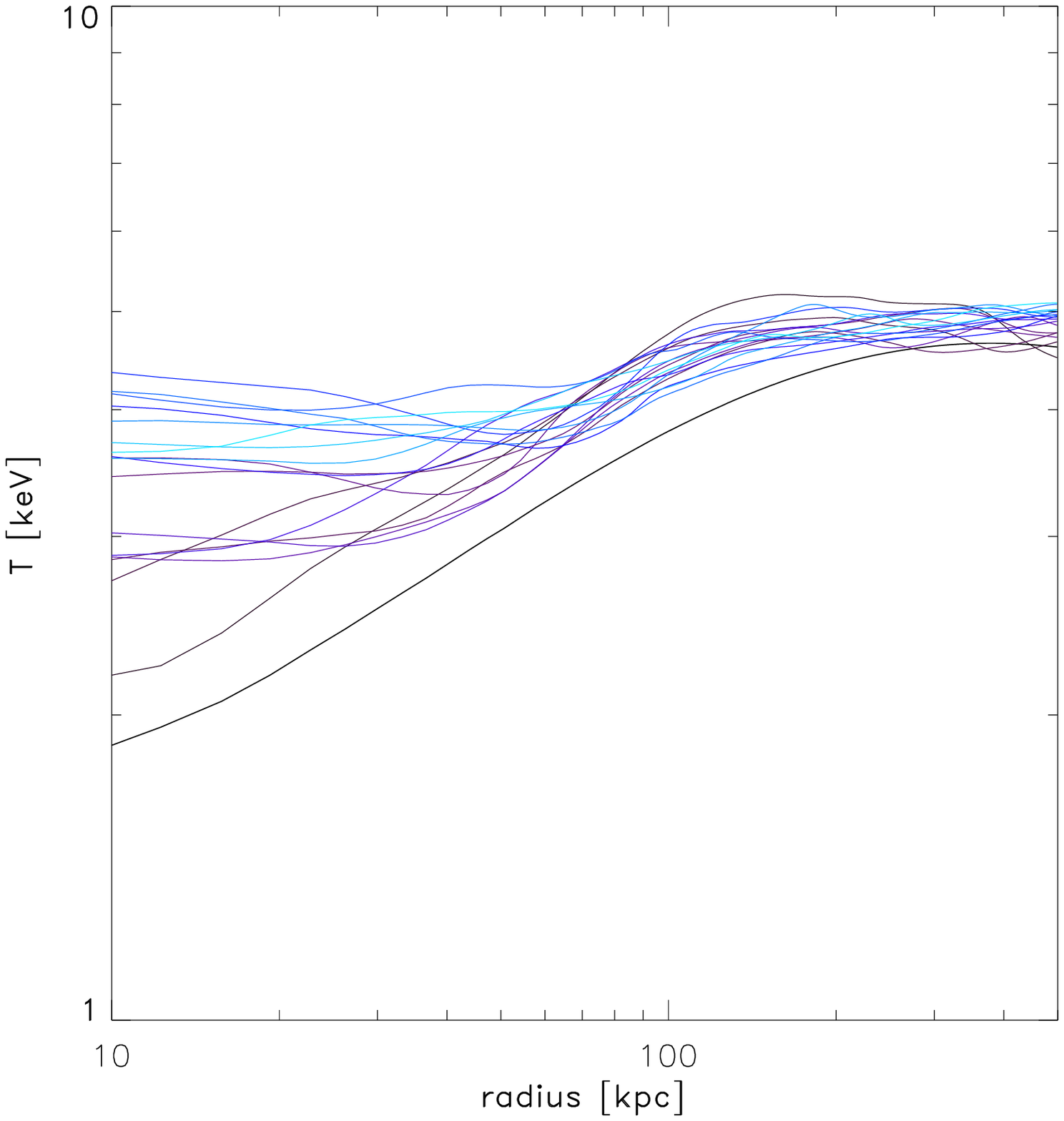} 
\caption{The evolution of temperature profiles for the strong heating models. 
The panels correspond to the the following pairs of parameters: 
(150,1.2), (200,0.9), (200,1.2), where the first number in the parenthesis
is the number of galaxies and the second is the galaxy mass in $10^{12}$ M$_{\odot}$, from left to right respectively.
The final time corresponds to 5 Gyr and the curves are plotted every 0.3 Gyr.}
\label{fig:temperature_strong}
\end{figure*}

In Figure \ref{fig:temperature_strong} we show the evolution of temperature profiles. This figure is for the models where heating is more efficient.
Specifically, it corresponds to the following pairs of parameters (150,1.2), (200,0.9), (200,1.2), where the first number in the parenthesis
is the number of galaxies and the second is the galaxy mass in $10^{12}$ M$_{\odot}$.
Progressively older profiles correspond to systematically brighter colors. The final time corresponds to 5 Gyr and the curves are plotted every 0.1 Gyr.
As is clearly seen in this figure, these models do not lead to the cooling catastrophe. Several features are of interest. The temperature profile does not asymptote toward an isothermal profile, as is generically the case when thermal conduction alone offsets cooling \citep{bregman88,guo08a,conroy08}. Despite the fact that we have not introduced an additional source of central heating such as an AGN, the cluster is able to remain in a thermally stable CC (i.e., with a central temperature which is lower than at the cooling radius) 
state via heat transport from the outer heater reservoir alone. Without fine-tuning, this is impossible to achieve with thermal conduction alone (when the cluster either becomes isothermal or undergoes a cooling catastrophe). Finally, the temperature profile is not always monotonic, but occasionally increases inward---a situation which is thermodynamically impossible if thermal conduction alone is operating. 
Note that these fluctuations are transient; such reversals are not present in the later stages of the evolution
(progressively lighter blue curves correspond to later times).
As we shall see in \S\ref{section:heating}, all of these features hint that an additional heat transport process is at play: turbulent heat diffusion.

\begin{figure*}
\includegraphics[width=0.33\textwidth]{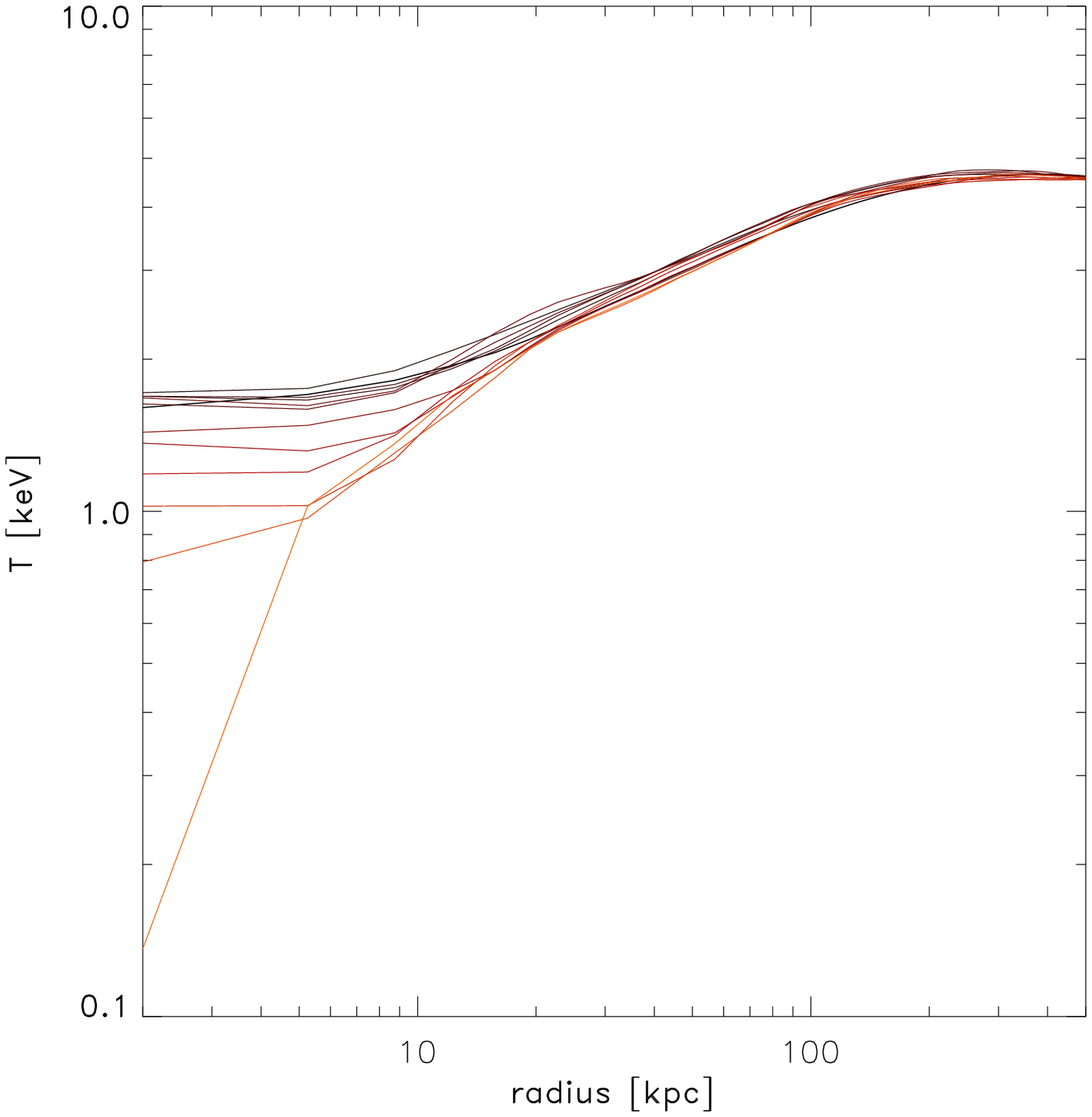}
\includegraphics[width=0.33\textwidth]{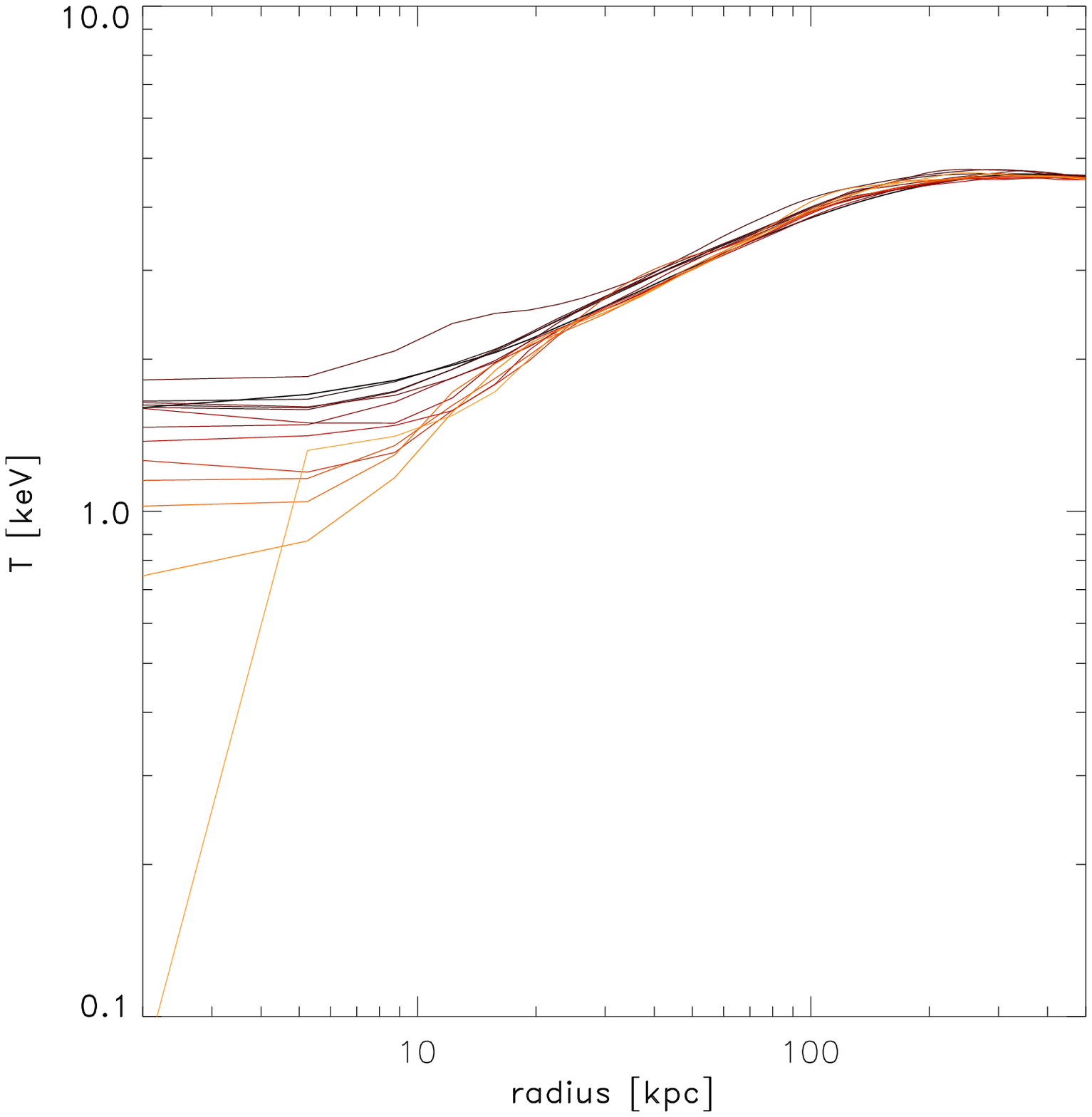}
\includegraphics[width=0.33\textwidth]{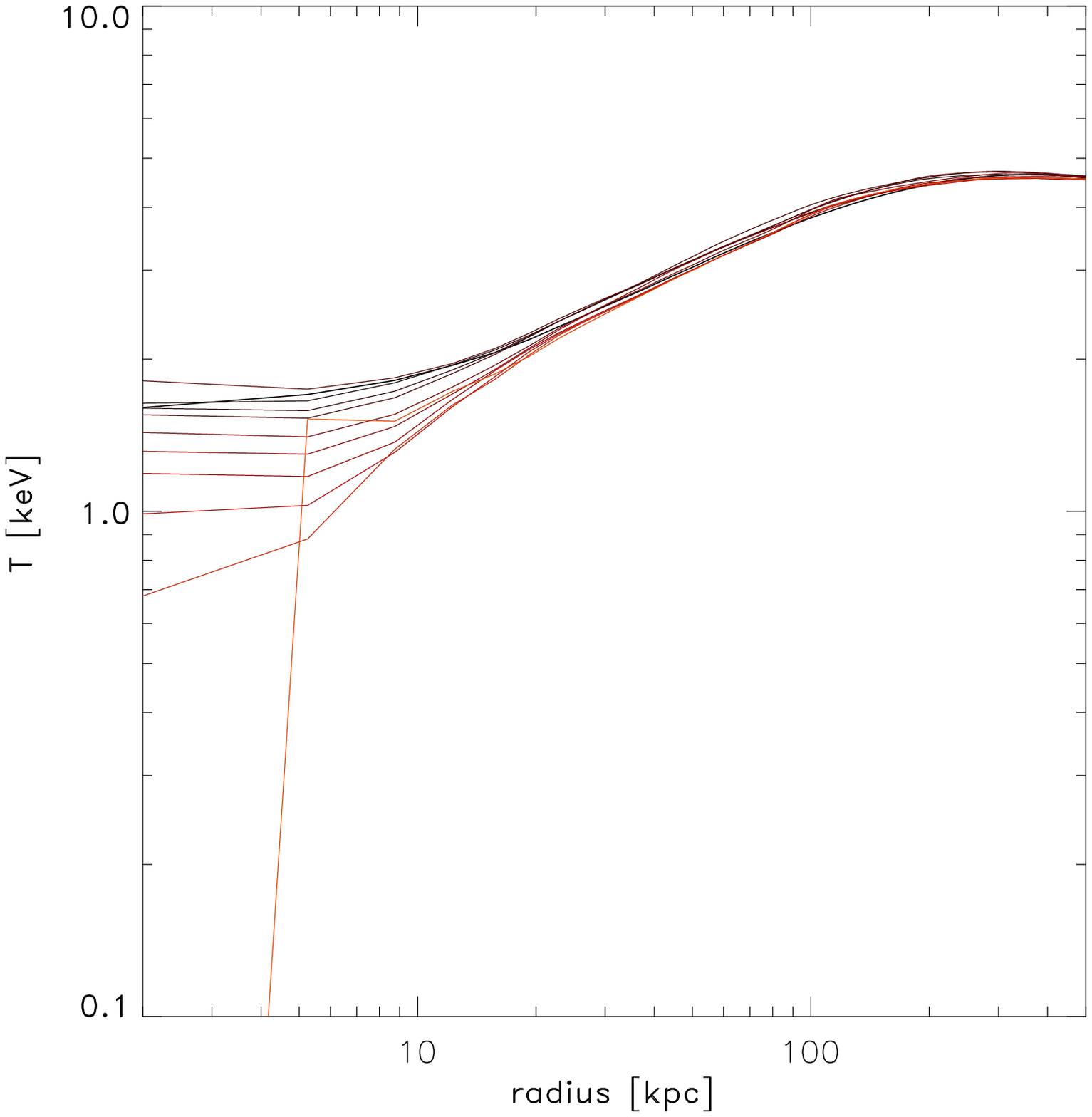} 
\caption{The evolution of temperature profiles for the weak heating models. 
From left to right are the results for the following sets of parameters: (100, 0.3), (50, 0.6), (50, 0.3),
where the first number in the parenthesis
is the number of galaxies and the second is the galaxy mass in $10^{12}$ M$_{\odot}$. The curves are shown
every 0.1 Gyr.}
\label{fig:temperature_weak}
\end{figure*}

Figure \ref{fig:temperature_weak} shows the temperature evolution for the parameters where the heating is least efficient. From left to right shown are: 
(100, 0.3), (50, 0.6), (50, 0.3). Here, the profiles are shown more frequently then in Figure 3 to better capture the evolution of the 
system just before the imminent cooling catastrophe. For the same reason, we also extend the radial scale to smaller distances from
the cluster center to show how the system becomes thermally unstable. It is evident that in all three cases, the cluster quickly evolves toward a
cooling catastrophe. In the final stages of the process, the cooling is so fast at the very center 
that the gas accretion accelerates so much that adiabatic compression in the shells surrounding the center can heat the gas up 
(e.g., see last profile in the right panel).

\begin{figure*}
\includegraphics[width=0.33\textwidth]{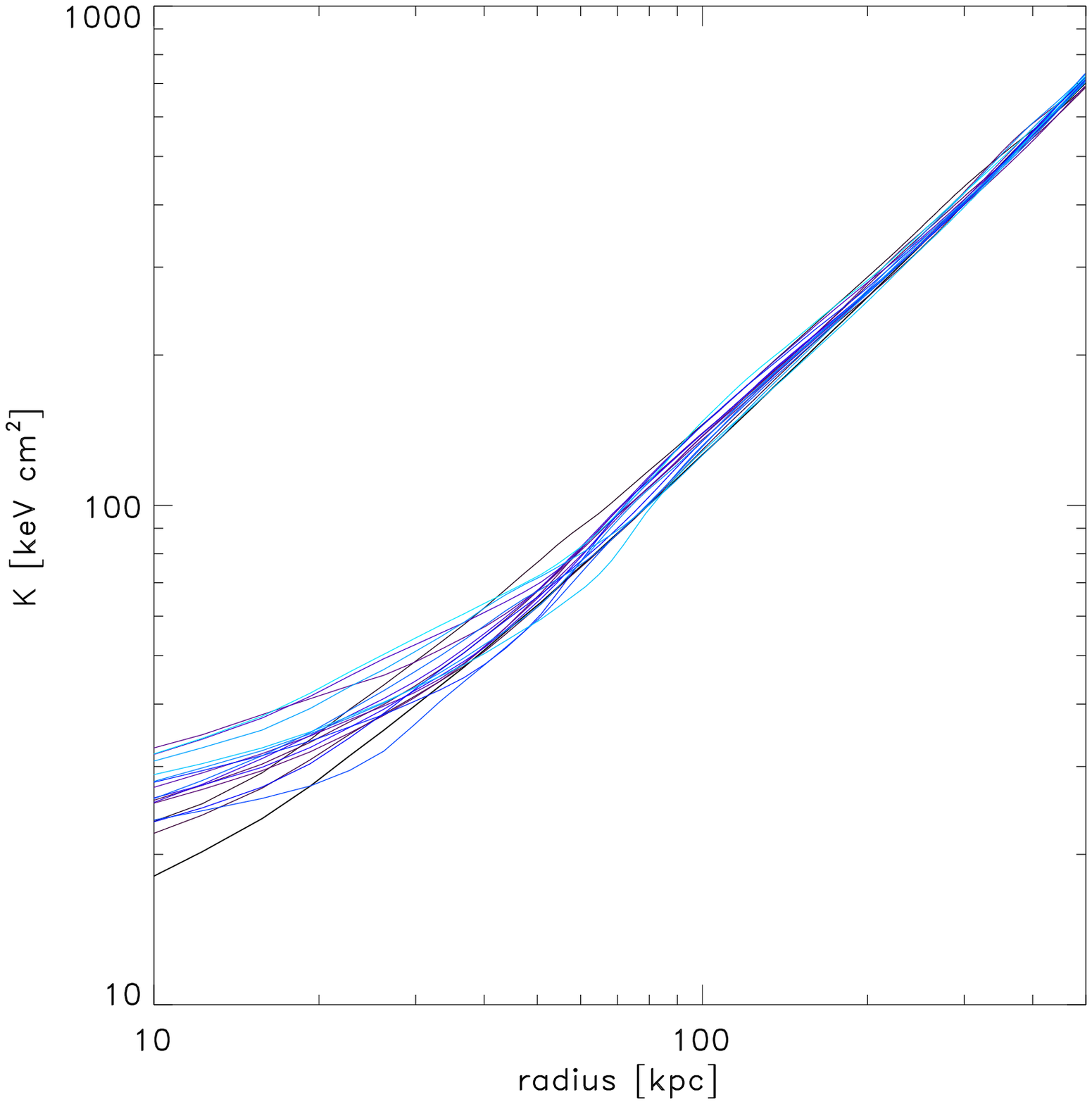}
\includegraphics[width=0.33\textwidth]{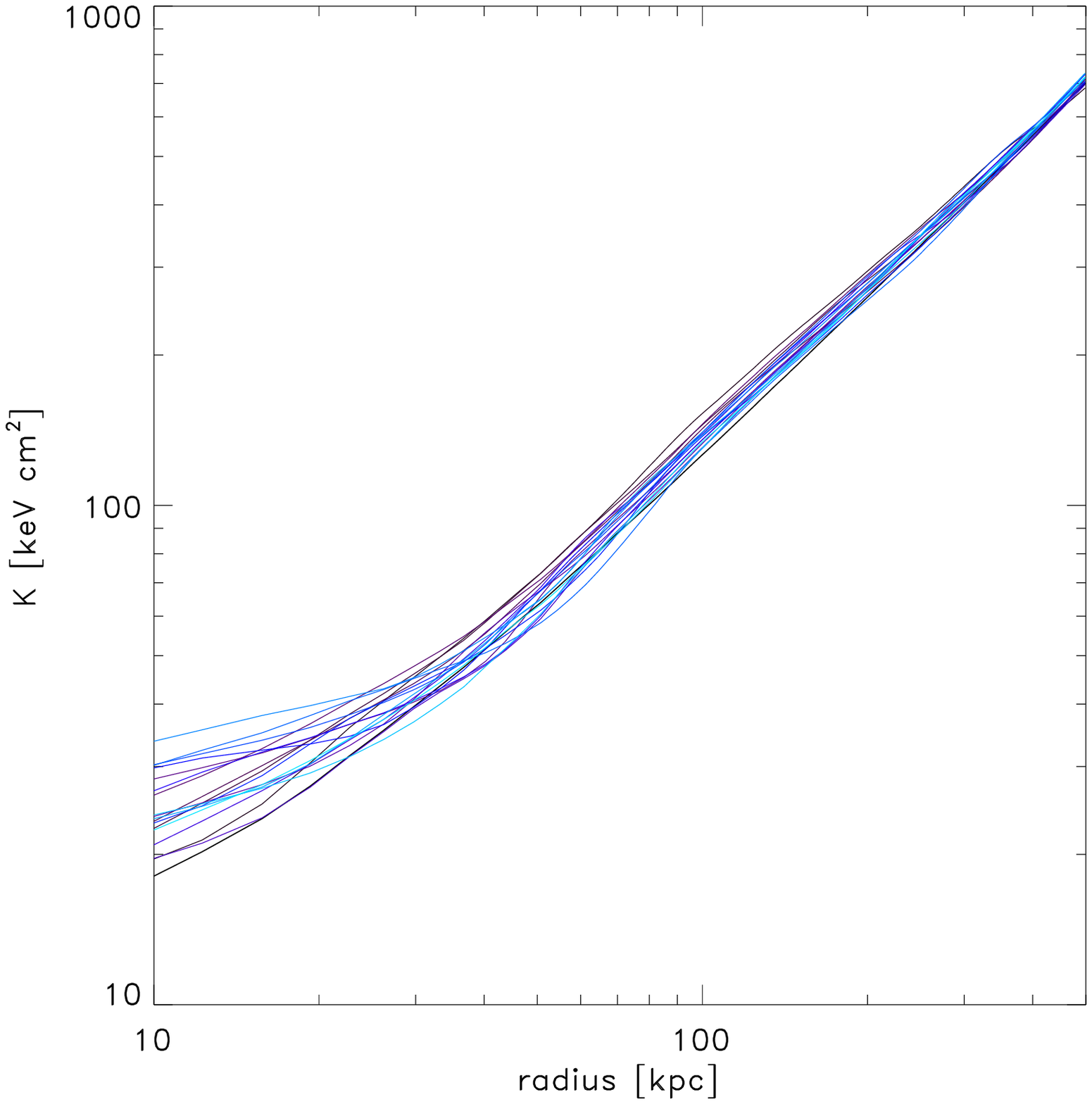}
\includegraphics[width=0.33\textwidth]{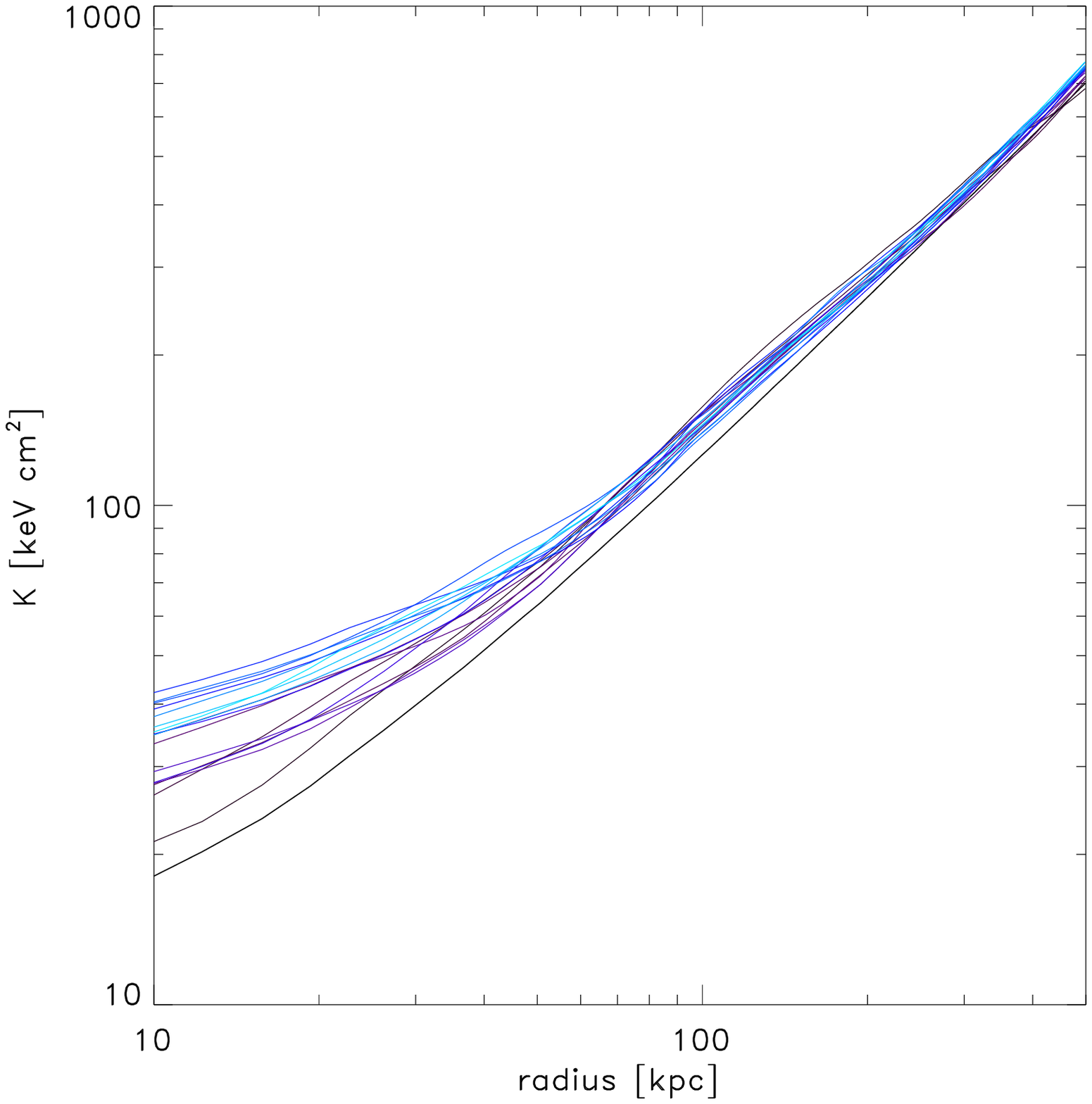} 
\caption{The evolution of entropy profiles for the strong heating models. 
From left to right are the results for the following sets of parameters: (150, 1.2), (200, 0.9), (200, 1.2),
where the first number in the parenthesis
is the number of galaxies and the second is the galaxy mass in $10^{12}$ M$_{\odot}$. The curves are shown
every 0.1 Gyr.}
\label{fig:entropy}
\end{figure*}

Finally, in Fig \ref{fig:entropy} we show the entropy profiles (where entropy is defined as $K\equiv k_{\rm B} T/n^{2/3}$) for the strong heating models. The central entropy grows somewhat, consistent with the rise in temperature, and as might be expected if heating by conduction and/or turbulent heat diffusion were taking place. However, these profiles show that turbulent convection/stirring is still a relatively gentle process; we do not see the flat isentropic central profile which might be expected if turbulent convection were extremely efficient. Instead, the fluid always remains stably stratified by entropy, which steadily increases outward at all times. 

\begin{figure}
\includegraphics[trim = 12mm 0mm 0mm 0mm, clip, width=0.45\textwidth, height=0.45\textwidth]{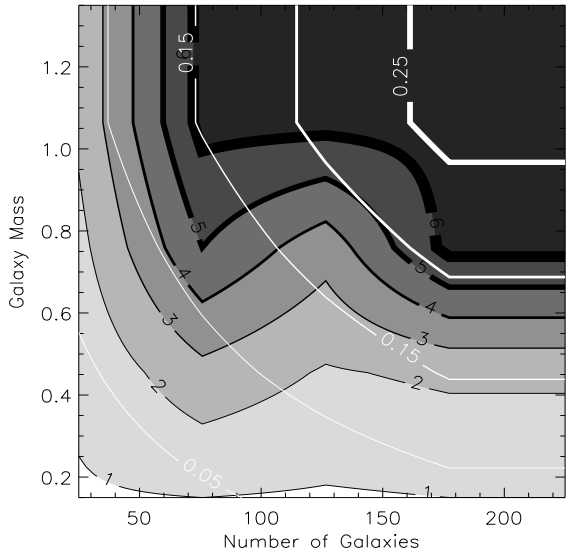}
\caption{Time until cooling catastrophe as a function of the number of galaxies and galaxy mass (in units of $10^{12}M_{\odot}$). 
Contours are in Gyr. Models that correspond to 6 Gyr (the maximum duration of the simulations) are 
thermally stable. See text for details.}
\label{fig:galaxy_parameters}
\end{figure}

\indent
As discussed above the cluster will develop a cooling catastrophe if the number of galaxies and/or their masses are too small. The time it takes for the
cluster to reach this point (essentially the effective cooling time) is plotted in Figure \ref{fig:galaxy_parameters} as a function of galaxy number and mass. 
The contours are plotted every Gyr. The models that exhibit the effective cooling time of 6 Gyr (the maximum simulation run time) are thermally stable. We point out that in practice
the models that possess cooling times $\ga 3$ to 4 Gyr could be considered stable as they are likely to experience cluster mergers that may reset
the conditions in the ICM and further slow down or essentially delay the cooling process. In any case, as can be seen in Figure 5, a substantial 
fraction of the models shows appreciably long effective cooling times. As a technical note, we add that the reason for the lack of monotonicity
in some of the contour lines as a function of galaxy number is that a single random seed was used to generate the conditions for a given number
of galaxies and varying galaxy masses.\\
\begin{figure*}
\includegraphics[width=0.45\textwidth]{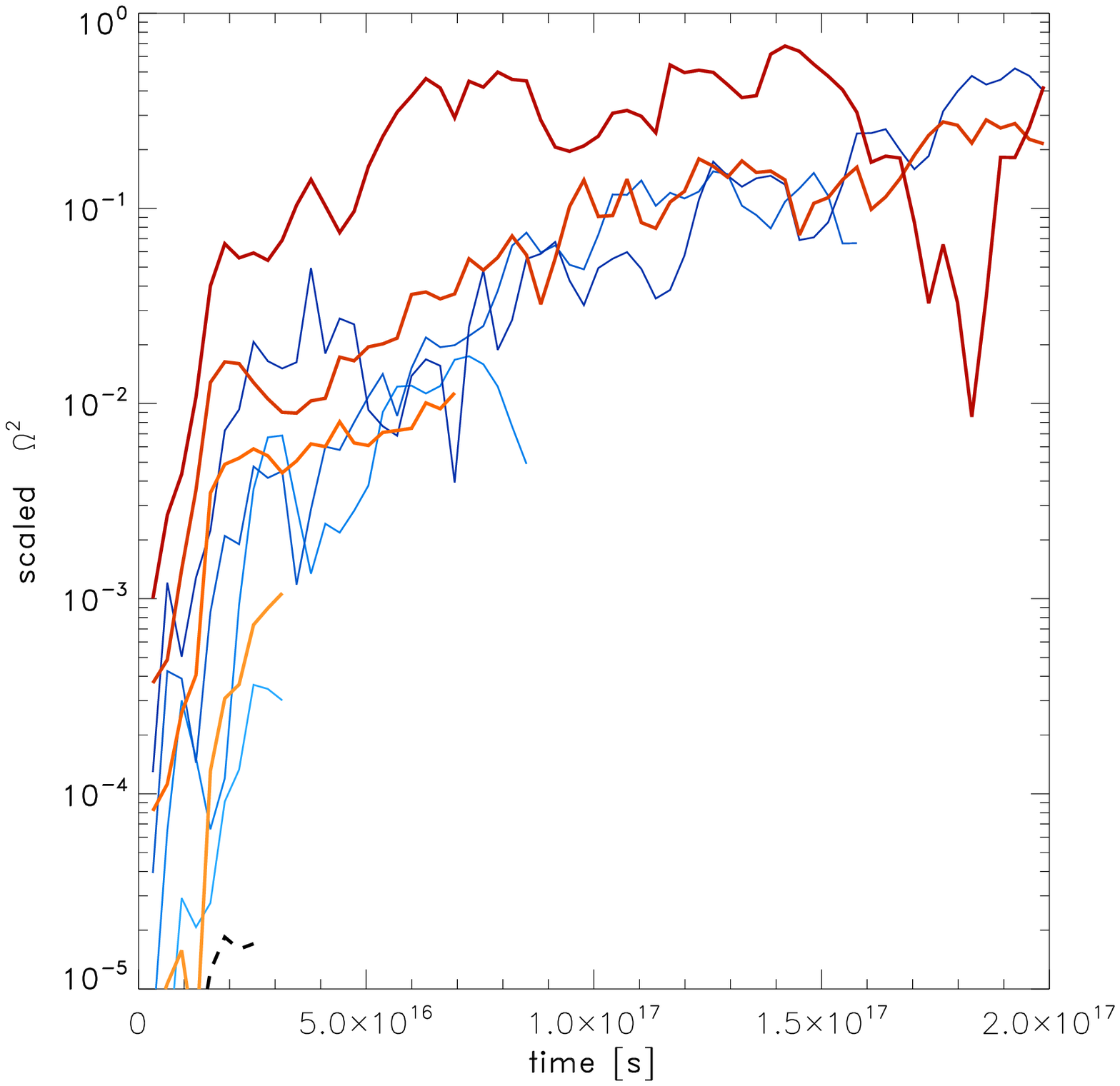}
\includegraphics[width=0.45\textwidth]{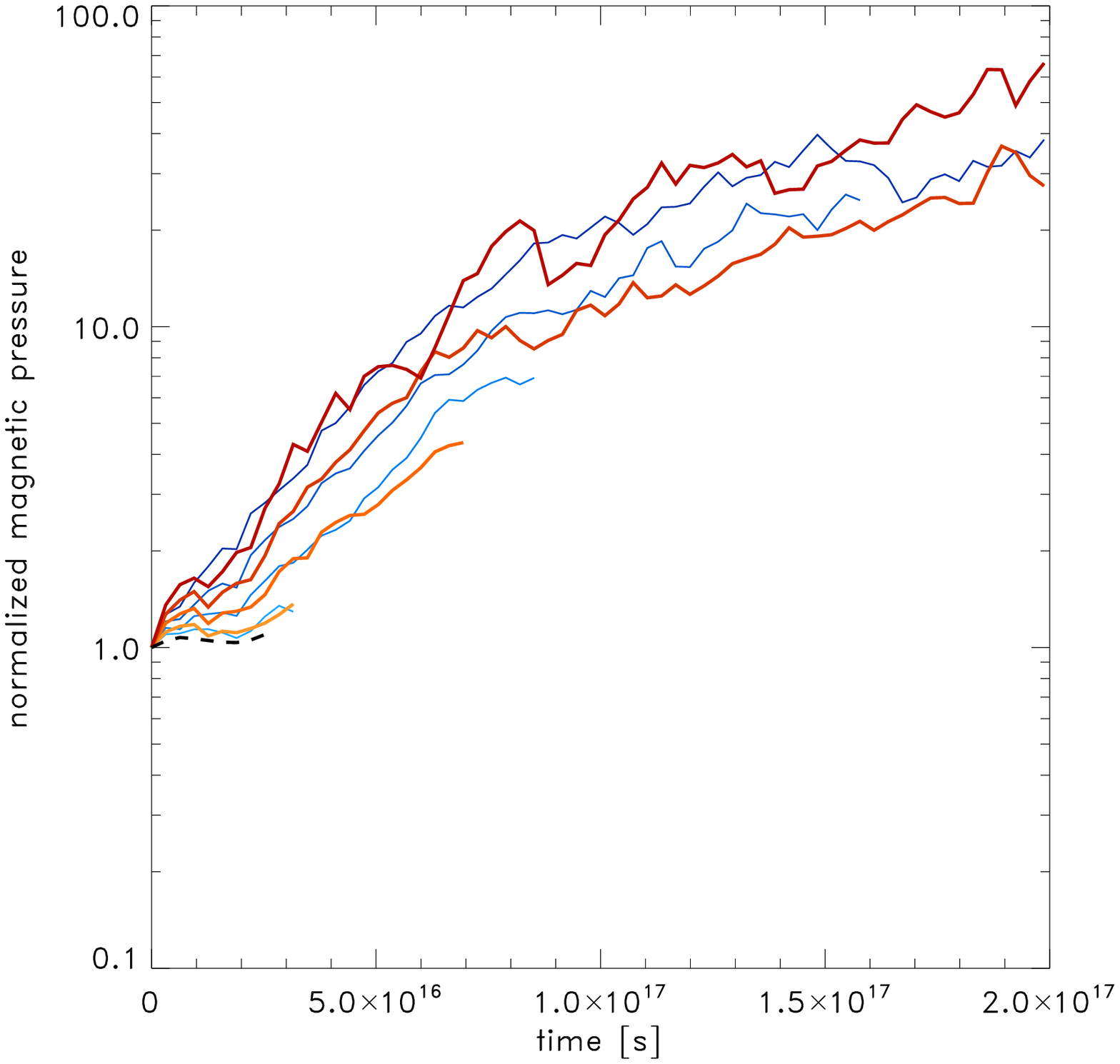}
\caption{The evolution of normalized vorticity (left panel) and normalized magnetic pressure. See text for definition of normalization. The curves correspond to the same dataset as that shown in 
Figure \ref{fig:velocities} and the meaning of lines is the same as in that figure.}
\label{fig:vort_B_field}
\end{figure*}

\subsection{Generation of vorticity and magnetic fields}
\label{section:vorticity}
As we have previously seen,  $g$-modes must be excited for stirring by galaxies to excite volume-filling turbulence. These $g$-modes also induce vorticity (equation \ref{eqn:vorticity}). Vorticity is therefore an excellent tracer of the growth of $g$-modes.   
We compute the evolution of vorticity in the central 100 kpc to assess if $g-$modes
are indeed generated and trapped. Figure \ref{fig:vort_B_field} (left panel) shows the evolution of the square of the 
scaled vorticity for the same set of parameters as in Figure \ref{fig:velocities} that shows velocity
dispersion and median velocity. The scaled vorticity is defined as 
${\bf \Omega}=(\lambda_{\rm ref}/\upsilon_{\rm ref})\nabla\times {\bf \upsilon}$, where $\lambda_{\rm ref}=50$ kpc and $\upsilon_{\rm ref}=100$ 
km s$^{-1}$ are the reference lengthscale and velocity, respectively. Thin blue lines are for 
100 galaxies and red ones for 200 galaxies. Galaxy gasses range from $3\times 10^{11}$ M$_{\odot}$ to $1.2\times 10^{12}$ M$_{\odot}$ and are 
uniformly sampled (lighter colors are for lighter galaxies). The black dashed line corresponds to the pure HBI case. A clear trend for
the vorticity to increase with time is seen in this figure, suggesting that $g-$modes are present and at least partially trapped, leading to the volume-filling turbulence seen. 

As discussed in \S\ref{section:theory}, a growth in vorticity might also lead to growth in the magnetic field; the possibility that magnetic fields could be turbulently amplified in clusters has been repeatedly raised (e.g., \citet{ruzmaikin89,subramanian06,ryu08,cho09}). Whilst a detailed study is beyond the scope of this paper, we check whether these theoretical expectations are satisfied in our simulations. Fig. \ref{fig:vort_B_field} shows that the magnetic energy density indeed grows in tandem with vorticity, with more vigorous stirring corresponding to greater field amplification. However, the characteristic growth time appears to be somewhat longer. Note that the simulations were initialized with extremely small magnetic fields: the initial plasma beta $\beta_{i} \gg \beta_{\rm obs}$, where  $\beta_{\rm obs} \sim 100$ is typically measured in the ICM. These small initial fields were for computational convenience (since the MHD approximation is satisfied with a trivially small magnetic field), and to ensure that magnetic fields never become dynamically important\footnote{Thus, as least in these simulations, the HBI is stabilized by the stirring motions and not by magnetic tension.}. Hence, despite growing by a factor of ~50, the magnetic energy density has not yet reached its saturated state, and is not yet in equipartition with turbulence. Nonetheless, the turbulent amplification of the B-fields, which mirrors the growth of vorticity, is a robust result.

\subsection{Relative contribution to gas heating}
\label{section:heating}

\begin{figure*}
\includegraphics[width=0.45\textwidth]{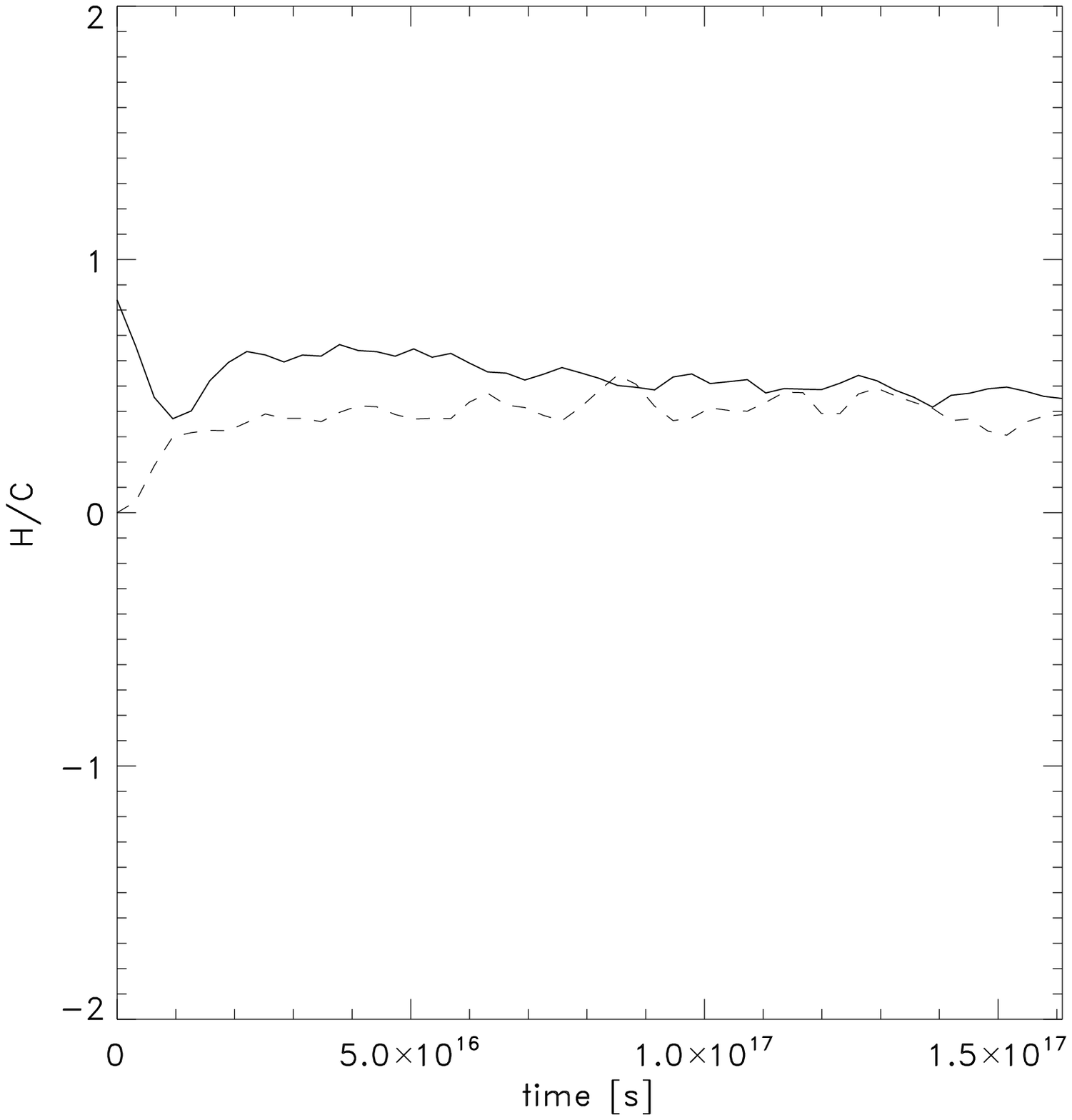}
\includegraphics[width=0.45\textwidth]{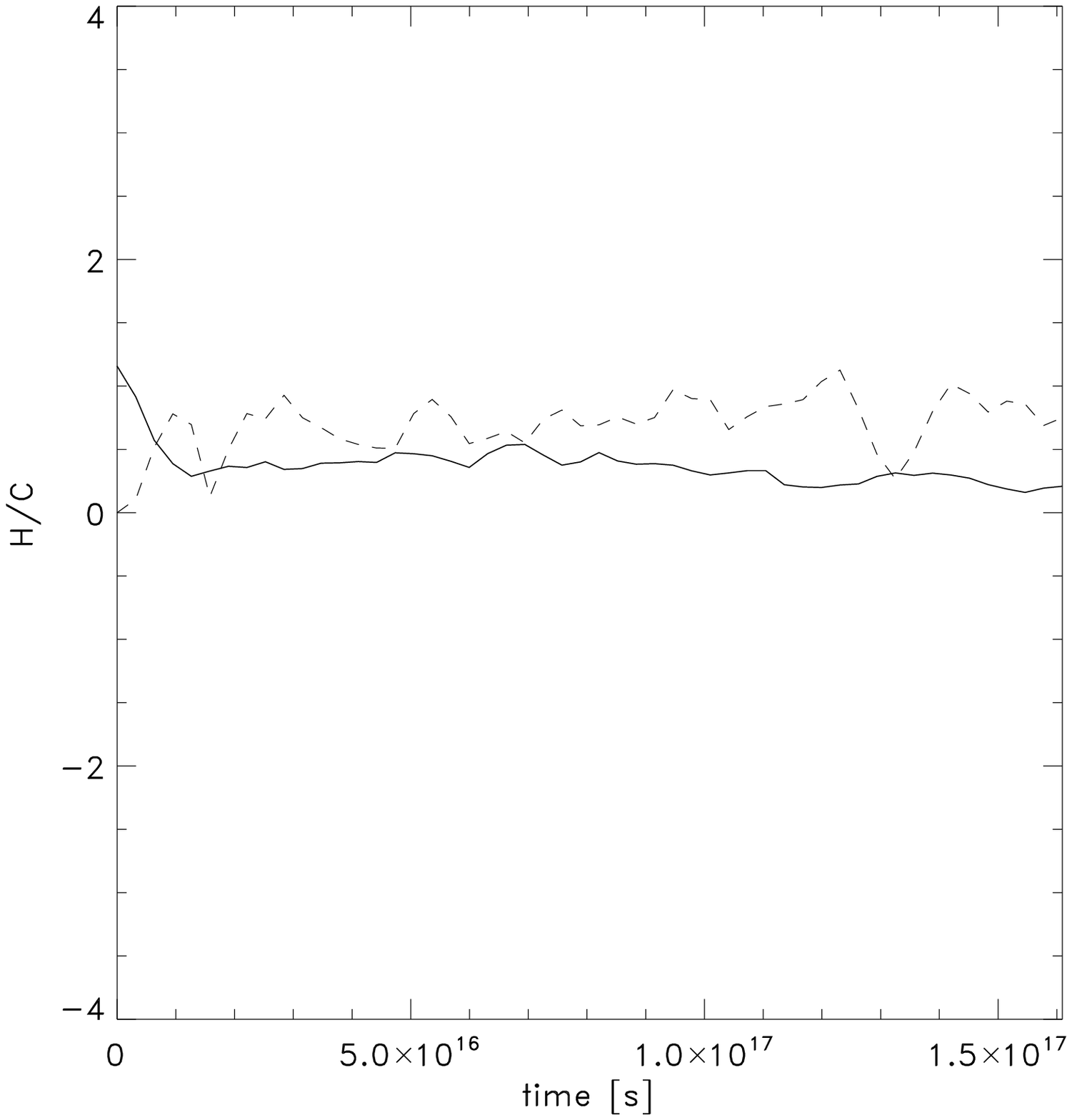}
\includegraphics[width=0.45\textwidth]{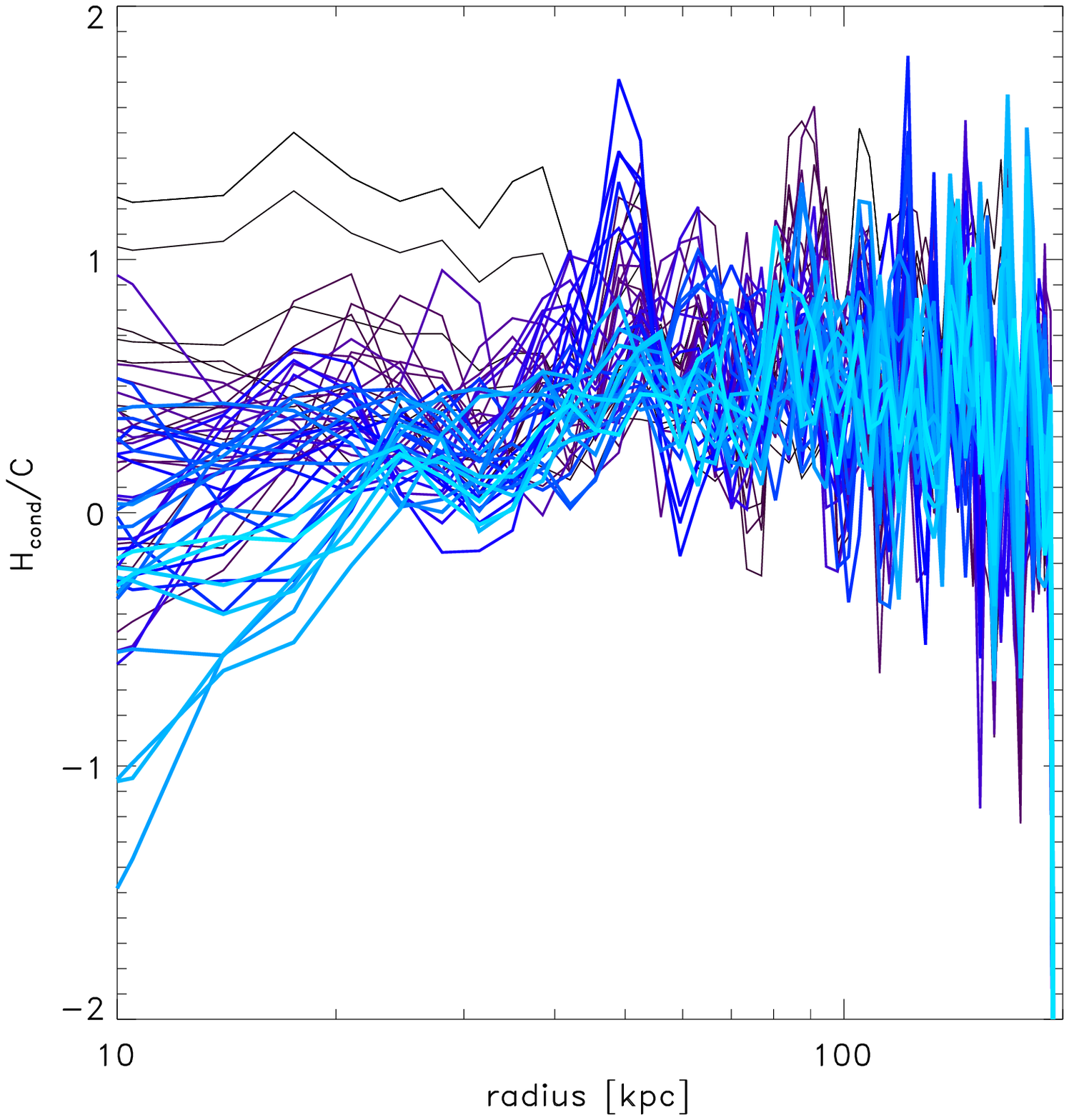}
\includegraphics[width=0.45\textwidth]{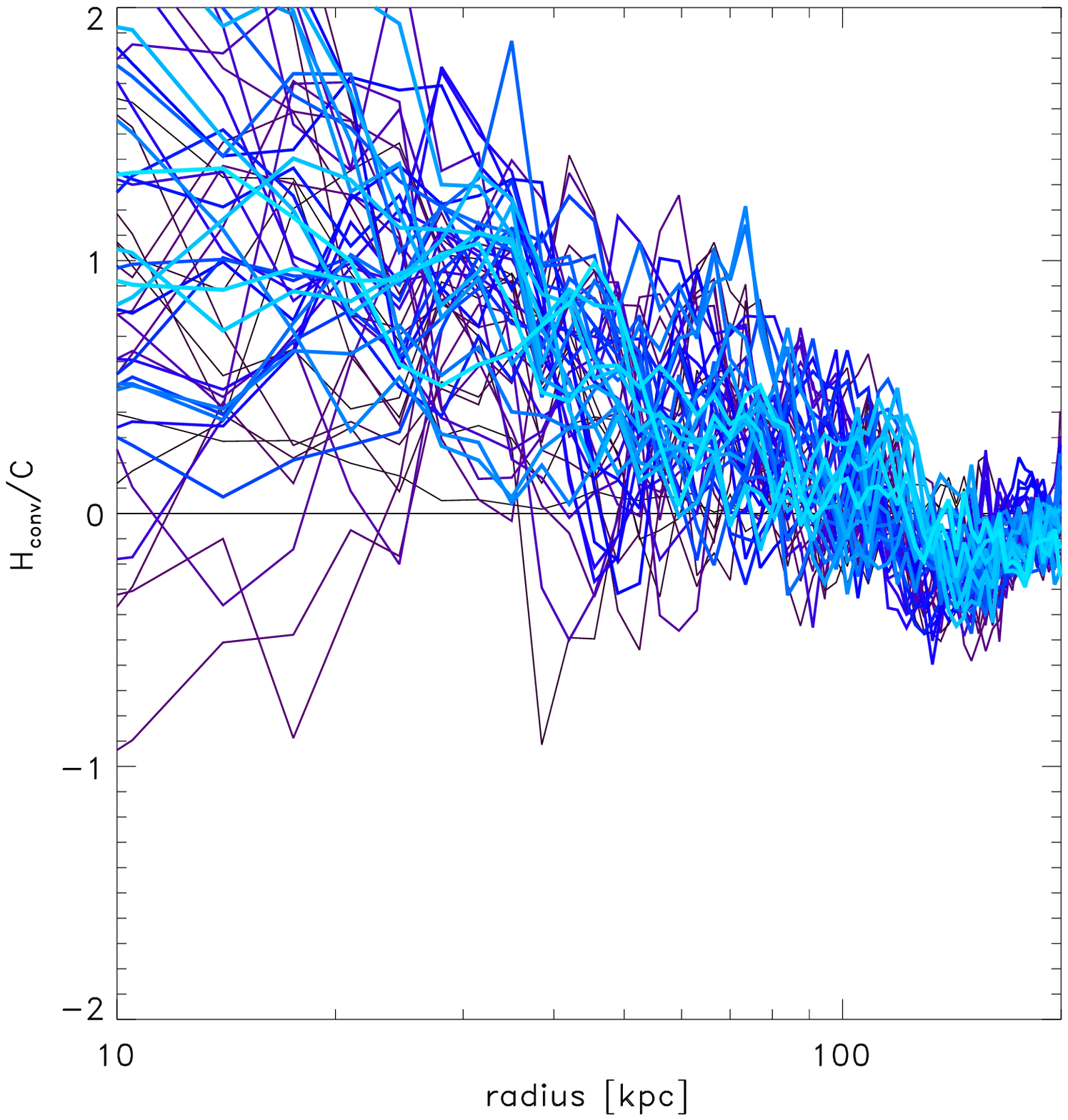}
\caption{Heating to cooling ratios for the case of steady volume-filling turbulence driven by 
a source function (see text and RO10 for more details). Top row: Conductive (solid line) and convective heating to cooling ratios as a function of time 
for the ICM within 100 kpc (left panel) 
and 50 kpc (right panel) from the cluster center, respectively.
Bottom row: Conductive (left panel) and convective (right panel) heating to cooling ratios as a function of radius. Progressively lighter blue color
denotes later times. The curves are plotted every $\sim100$ Myr.}
\label{fig:heat_cool_strong}
\end{figure*}

\begin{figure*}
\includegraphics[width=0.45\textwidth]{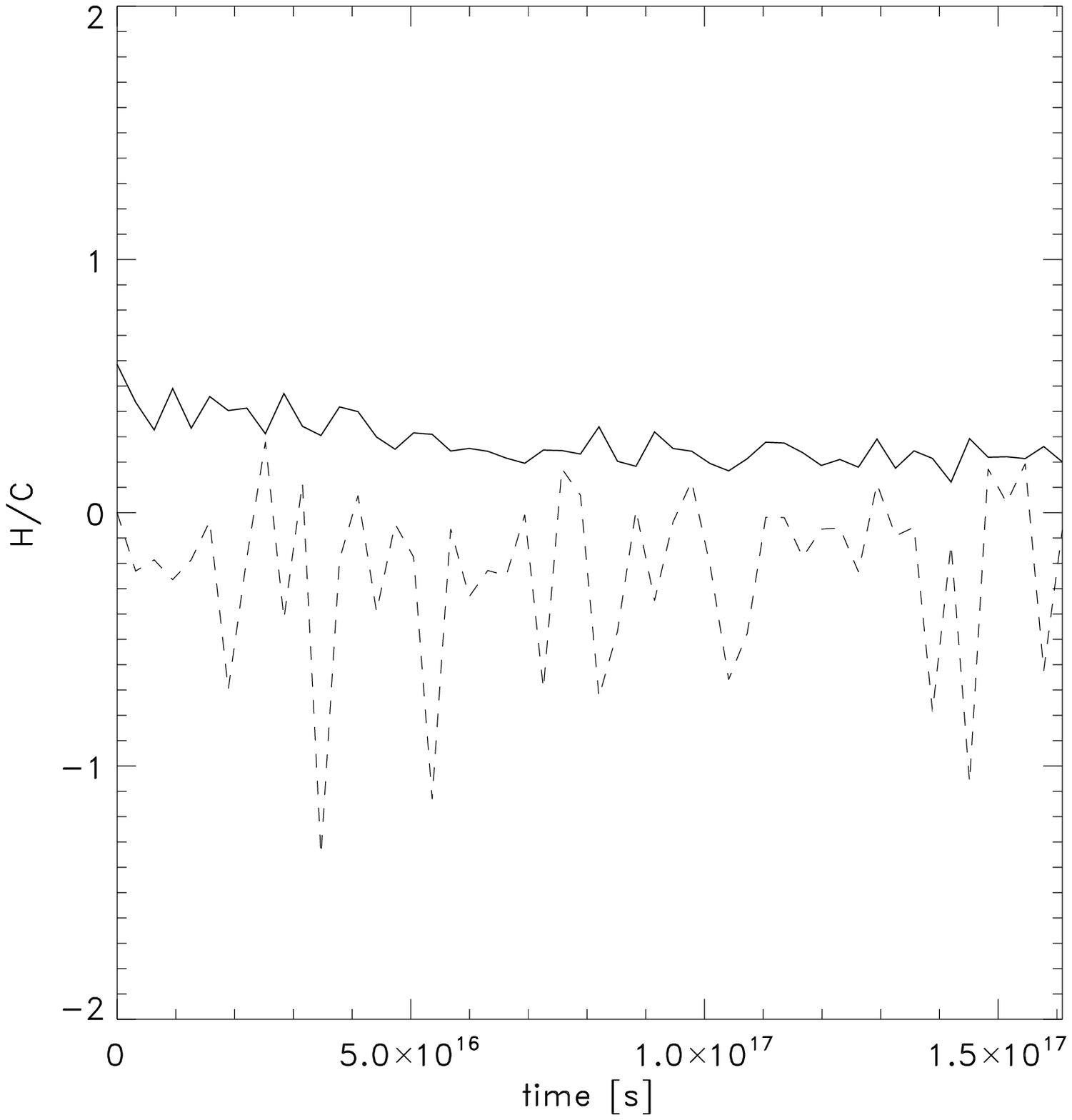}
\includegraphics[width=0.45\textwidth]{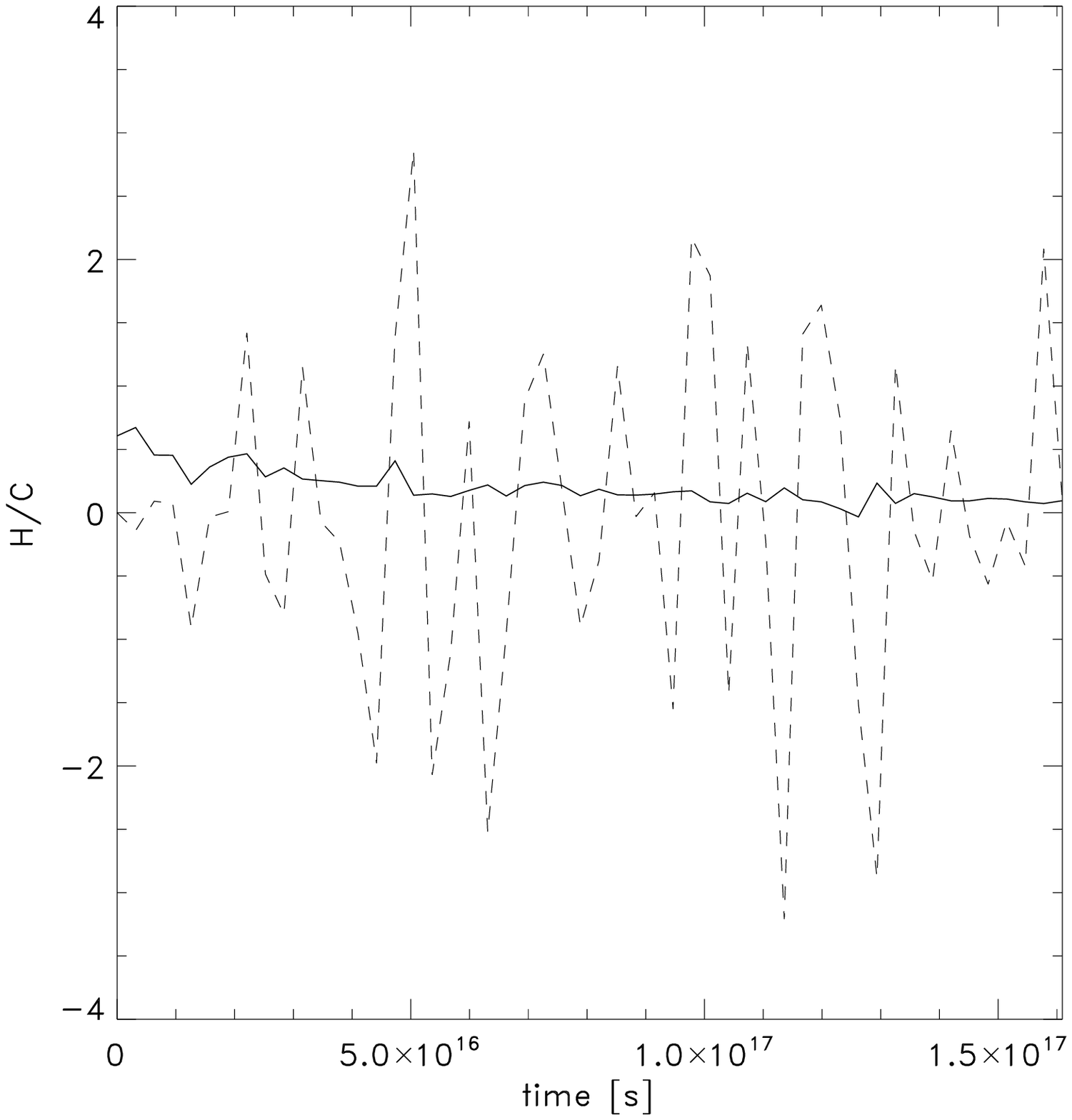}
\includegraphics[width=0.45\textwidth]{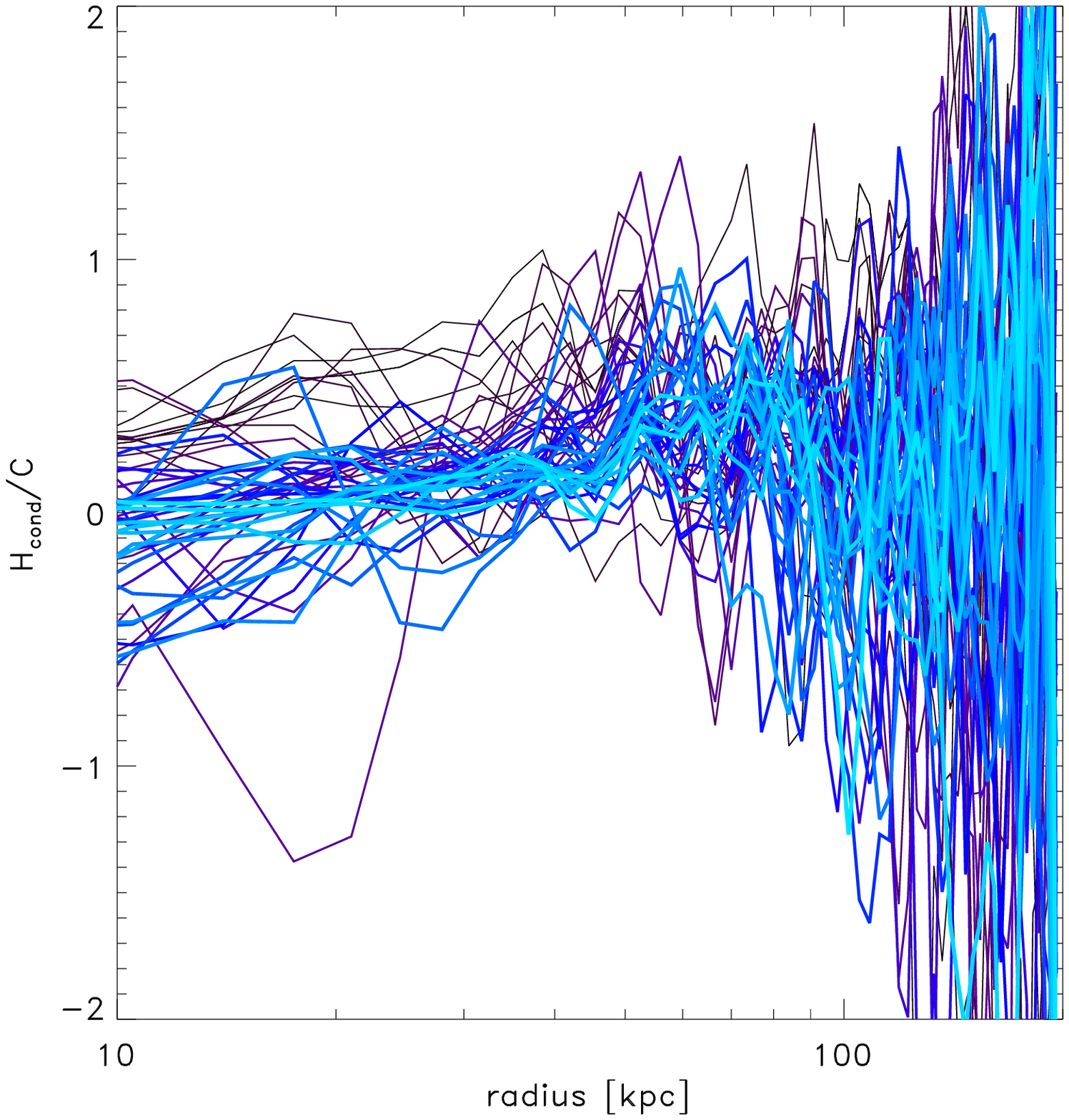}
\includegraphics[width=0.45\textwidth]{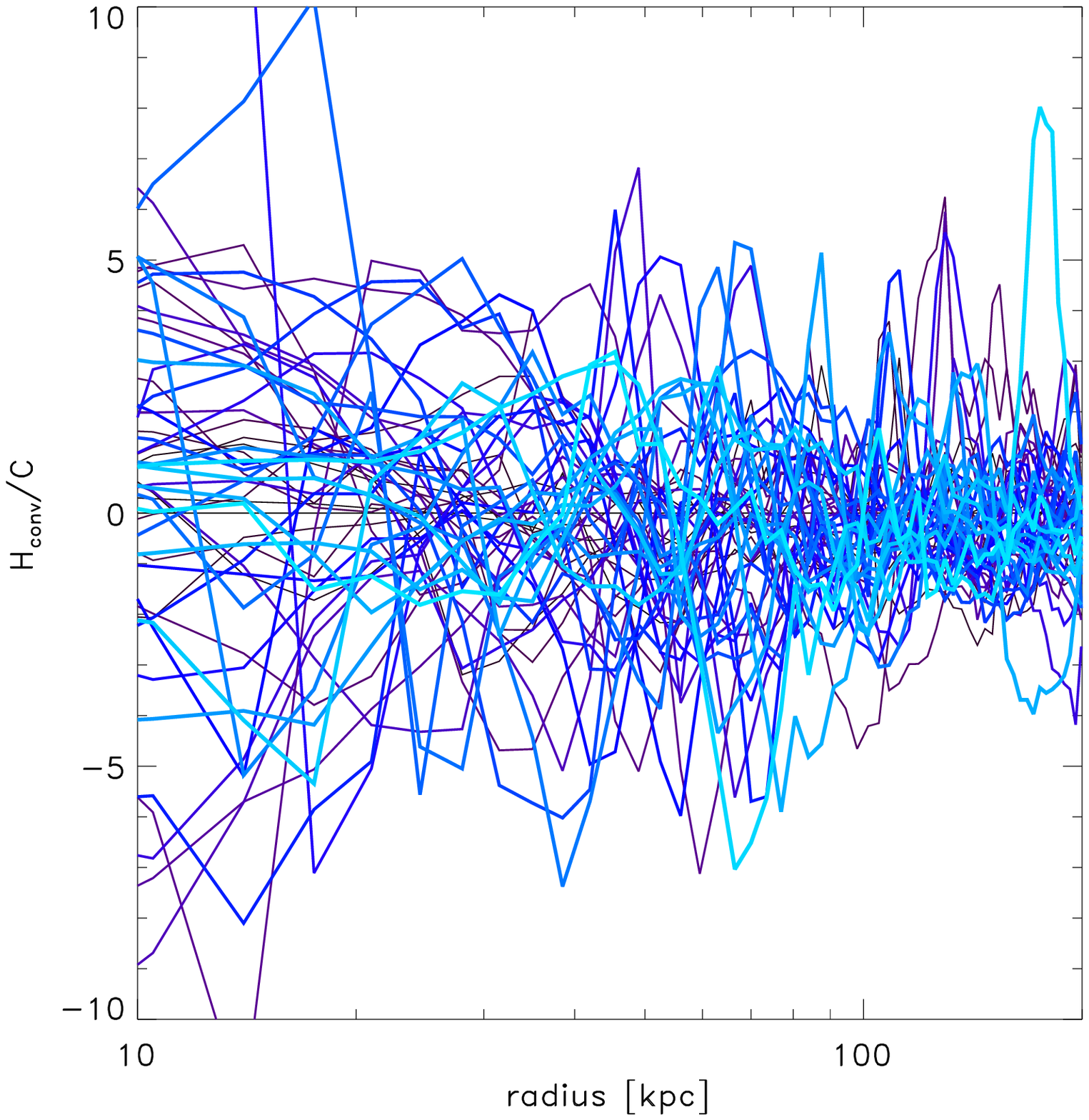}
\caption{Same as for Fig \ref{fig:heat_cool_strong} for the model of turbulence stirred by galaxy motions. 
Top row: Conductive (solid line) and convective heating (dashed line) to cooling ratios as a function of time 
for the ICM within 100 kpc (left panel) 
and 50 kpc (right panel) from the cluster center, respectively.
Bottom row: Conductive (left panel) and convective (right panel) heating to cooling ratios as a function of radius. The curves are plotted every $\sim100$ Myr; progressively lighter blue color
denotes later times.}
\label{fig:heat_cool_galaxies}
\end{figure*}

In \S\ref{section:temperature} we noted a number of interesting features in the temperature profiles of our stable clusters. They remained stable CC clusters, neither becoming isothermal nor developing cooling catastrophes, as clusters stabilized solely by thermal conduction generally do. Furthermore, the central temperature showed time-dependent oscillations, sometimes becoming hotter than gas further out. A temperature inversion would not happen if only thermal conduction was at play. This behooves us to take a closer look at what actually stabilizes the customary thermal runaway. We have already discussed the effect that turbulence can have on thermal conduction, by tangling field lines and countering the HBI. However, turbulence itself can be a source of heating, either via viscous dissipation of turbulence, or turbulence diffusion of high entropy gas into low entropy regions (e.g., \citet{dennis05}, and references therein). Let us examine these in turn. 

As long as there is sufficient separation of scales that an inertial range can develop (such that the energy per unit mass per unit time $\epsilon \sim v^{3}/l$ is independent of scale), the heating rate from dissipation of turbulent motions is independent of the nature of viscosity. In particular, it is unimportant if our numerical viscosity is different from the actual physical viscosity in the ICM. The heating rate per unit volume due to dissipation of such motions is \citep{dennis05}: 
\begin{equation}
\Gamma = \frac{c_{\rm diss} \rho v_{\rm t}^{3}}{l} =  \frac{c_{\rm diss} U_{\rm t}}{t_{\rm edd}}
\end{equation}
where $l$ is the dominant lengthscale, $U_{t}$ is the energy density in turbulence, and $t_{\rm edd} = l/v_{t}$ is the eddy-turnover time on the dominant lengthscale. To estimate $t_{\rm edd}$, we can note that vorticity $\Omega = \nabla \times v_{t}$ has units of $t_{\rm edd}^{-1}$, and that our scaled vorticity in Fig \ref{fig:vort_B_field} is $\Omega_{s}^{2} \approx 0.1 (\lambda_{\rm ref}/50 \, {\rm kpc})(v_{\rm ref}/100 \, {\rm km \, s^{-1}})^{-1}$. This implies
\begin{equation}
t_{\rm eddy} \approx 1.5 \times 10^{9} \left( \frac{\Omega_{s}}{0.3} \right)^{-1} \, {\rm yr}.  
\end{equation}
Consistently, note that the vorticity in Fig \ref{fig:vort_B_field} indeed takes $\sim 1$ Gyr to rise to its asymptotic value. This implies that the heating time for turbulent dissipation of motions is: 
\begin{equation}
t_{\rm heat} = \frac{\rm U_{\rm thermal}}{\Gamma} = c_{\rm diss} \left( \frac{3}{\gamma} \right) {\mathcal M}^{2} \, t_{\rm eddy} \sim 10^{11} \, {\rm yr} 
\end{equation} 
where ${\rm U_{t}}$ is the thermal energy density, and we have defined the turbulent Mach number ${\mathcal M} \equiv v_{t}/c_{s}$ (note that our quoted velocities $v_{t}$ are in 3D). While there are factors of order unity uncertainty, it is clear that the mild subsonic motions we explore are a negligible source of heating via viscous dissipation (and consistent with other estimates; \citet{dennis05}). This also implies that dynamical friction heating due to galaxy motions \citep{elzant04,kim05,kim07,conroy08,birnboim10} is not the source of heating which averts the cooling catastrophe in these simulations.

On the other hand, turbulent heat diffusion is not negligible. One can estimate its contribution from a simple mixing length prescription as in equation \ref{eqn:heat_diffusion}; this shows that it can be at least comparable to and may exceed the thermal conduction contribution. However, the coefficient of turbulent diffusivity, $\kappa_{\rm turb} \sim u l$, is only approximate and subject to order unity corrections. Since we have full knowledge of the density, velocity, temperature and magnetic fields in our simulations, we can attempt to directly compute the heating contributions from thermal conduction and turbulent heat convection. In particular, at a radius $r$ we can calculate the inward heat flux due to conduction:
\begin{eqnarray}
{\bf F}_{\rm cond}=-\kappa \hat{\bf e}_{B}(\hat{\bf e}_{B}\cdot{\mathbf\nabla} T),\\ \nonumber
\end{eqnarray}
\noindent
where $\hat{\bf e}_{B}$ is a unit vector pointing in the direction of the magnetic field and $\kappa$ is the Spitzer-Braginskii 
conduction coefficient given by $\kappa = 4.6\times 10^{-7}T^{5/2}$erg s$^{-1}$cm$^{-1}$K$^{-1}$, as well as the convective heat flux \citep{parrish08}:
\begin{equation}
{\bf F}_{\rm conv} = \frac{\gamma}{\gamma-1}k_{\rm B} (\langle v \rangle \langle \delta n \delta T \rangle + \langle n \rangle \langle \delta v \delta T \rangle ) + ( \langle \delta n \delta T \delta v \rangle ) 
\label{eqn:convection} 
\end{equation}  
where $\langle x \rangle$ is the spatial average of quantity $x$ in the shell and $\delta x$ is the local deviation of that quantity from its average; generally the second term is dominant. We can then compare these to the total rate of energy loss within radius $r$ due to radiative cooling. We can also compute the volumetric heating rate due to these two processes, via $H_{\rm cond} = \nabla \cdot {\bf F}_{\rm cond}$, $H_{\rm conv} = \nabla \cdot {\bf F}_{\rm conv}$, although these are of course much noisier quantities.  

It is useful perhaps to begin by considering a case where the properties of turbulence are well known: the 'strong' driven turbulence case of RO10, which has volume filling turbulence by construction, and rms velocities of $\sim 150 {\rm km \, s^{-1}}$. Conductive (solid line) and convective (dashed line) heating to cooling ratios as a function of time 
for the ICM within 100 (50) kpc from the cluster center are shown in the upper left (right) panel of Fig \ref{fig:heat_cool_strong}. It is clear that conduction only contributes $\sim 50\%$ of the heat necessary to overcome cooling and convective heat flow is an important part of the energy budget; indeed, in the central regions turbulent advection of heat is the dominant heating process (note that the cluster is not in complete equilibrium, so the sum of the two ratios is not necessarily unity). The convective heat flow shows time-dependent fluctuations, as might be expected. In the bottom panels, we show the volumetric heating to cooling ratios as a function of radius, for conduction (left panel), and turbulent convection (right panel). The curves are plotted every ~100 Myr; progressively lighter colors denote later times. The divergence of heat fluxes is a much noisier quantity, as reflected in the plots. Nonetheless, it is clear from the plots that convective heating dominates near the center, whilst conductive heating dominates further out. This reminiscent of stable hybrid AGN+conduction heating models \citep{ruszkowski02} where the AGN heats the cluster center and conduction is important further out.  

In Fig. \ref{fig:heat_cool_galaxies}, we show the same plots, but for the case where turbulence is due to stirring by galaxies. All results presented in this figure are for the case of 200 galaxies, each with $9\times 10^{11}$ M$_{\odot}$; this is stable against a cooling catastrophe. As before, conduction is only a fraction $\sim 30-50\%$ of the energy budget. However, in this case convective heating shows dramatic oscillations as a function of time; the amplitude of the oscillations $H_{\rm conv}/C \sim 5-10$ near the center is much larger than in the driven turbulence case $H_{\rm conv}/C \sim 1-2$. The reason for this is that the dominant lengthscales of motion are comparable or larger than the depicted radii, as might be expected if $g$-modes are excited (since most of the energy in $g$-modes are in the largest lengthscales, comparable to the trapping radius). For instance, from \S\ref{section:vorticity}, a typical lengthscale on which vorticity is excited is $\lambda \sim v_{t} |\Omega|^{-1} \sim 150 \, {\rm kpc} \, (v_{t}/100 \, {\rm km \, s^{-1}}) (\Omega_{s}/0.3)^{-1}$. Since fluctuations in the velocity field span larger scales than the ones under interest, our calculation of $H_{\rm conv}$ will show strong time dependence (however, the calculations of the conductive heat flux are of course still valid. Note that $H_{\rm conv}/C+H_{\rm cond}/C$ has to be unity on average, since the cluster is stabilized against a cooling catastrophe). As noted earlier, Poisson fluctuations in the number of galaxies in the core will also drive time-dependent fluctuations in the velocity field. The gas is sloshing in the potential well; we observe this directly too in the simulations, as the gas pressure maximum wanders in time from the center of the potential well. Nevertheless, despite the breakdown of equation \ref{eqn:convection} in a rigorous sense, it is clear from the amplitude of fluctuations in the bottom panels of Fig \ref{fig:heat_cool_galaxies} that (as in the driven turbulence case) conductive heating increases outwards in radius, while convective heating is more important near the center. In particular, the dominance of convective heating near the center, and its positive and negative fluctuations, allow us to understand the fluctuating temperature profiles seen in Fig \ref{fig:temperature_strong}. Since conduction is only a part of the energy budget, there is no reason for the stabilized temperature profile to approach isothermality. Furthermore, the reason why the central temperature gradient can occasionally become inverted (with the center hotter than its surroundings) is clear: if a high-entropy fluid element is compressed at the center, this will result in higher central temperatures. While thermal conduction seeks to make the fluid isothermal (since heat flows down the temperature gradient), turbulent heat diffusion seeks to make the fluid isentropic (since heat flows down the entropy gradient). In this sense, the subsonic turbulence induced by galaxies results in only mild convection, since as seen in Fig \ref{fig:entropy}, the gas remains convectively stable with entropy increasing monotonically outward. Whilst we have not directly calculated the diffusion of metals directly, this also suggests that metal mixing to larger radii will be somewhat enhanced (so that metals will have a broader distribution than the galaxies), but not greatly so. Indeed, a mixing-length theory calculation of metal dispersal via turbulent diffusion by \citet{rebusco05}, who assumes levels of turbulence very similar to those we have simulated, shows excellent agreement with observations.  

\begin{figure}
\includegraphics[width=0.45\textwidth, height = 0.45\textwidth]{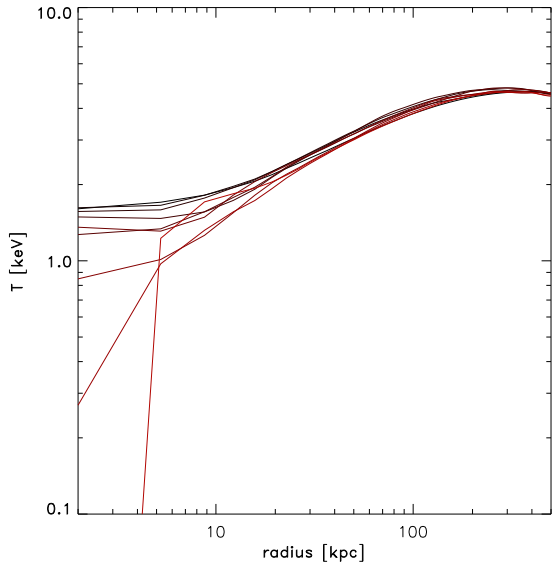}
\caption{Evolution of temperature profiles for a purely hydrodynamic run (no thermal conduction) with 200 galaxies of mass $1.2 \times 10^{12} \, {\rm M_{\odot}}$; our strongest heating model. Curves are shown every 0.1 Gyr. The cluster rapidly undergoes a cooling catastrophe, demonstrating that thermal conduction is an essential element in stabilizing the cluster; purely hydrodynamic turbulent heat diffusion alone is insufficient.}
\label{fig:no_conduction}
\end{figure}

Could turbulent heat diffusion alone stabilize a thermal runaway? We tested this hypothesis by running purely hydrodynamic simulations of our most vigorous stirring case (200 galaxies of mass $1.2 \times 10^{12} \, {\rm M_{\odot}}$), without thermal conduction. The results are shown in Fig. \ref{fig:no_conduction}. The cluster rapidly undergoes a cooling catastrophe (consistent with the results of \citet{kim07}), demonstrating that thermal conduction is an essential element in stabilizing the cluster. Note that this model already represents the most extreme choice in parameter space of the amount of stirring possible by galaxies. Also in line with these results, purely hydrodynamic simulations of 'sloshing'\citet{zuhone10} show that the cooling catastrophe can be delayed by not disrupted. Besides providing the dominant source of heating in the outer regions, thermal conduction also reduces stabilizing buoyancy forces (as discussed in \S\ref{section:theory}) and thus enables more rapid, efficient mixing and turbulent heat diffusion. Passive scalars such as metals are more efficiently advected in the presence of conduction \citep{sharma09}, and the same is likely true of the advection of entropy. Thus, intriguingly, whilst neither process alone can stabilize the cluster, the interplay between turbulence and conduction does permit stability: turbulence enables conduction, and conduction enables turbulence. 

\section{Conclusions}
\label{section:conclusions}

Using three-dimensional MHD simulations, we have studied 
the effect of anisotropic thermal conduction and stirring motions due to galaxies orbiting in the cluster potential
on the effective cooling rate in cluster cool cores. Such galaxies excite mild subsonic turbulence with $v_{\rm t} \sim 100-200 \, {\rm km \, s^{-1}}$. We find that a combination of thermal conduction and turbulent heat transport can stabilize the cluster, for realistic parameter choices consistent with gravitational lensing observations of substructure in clusters. Unlike much previous work, there is no subgrid physics in our simulations: we do not invoke sub-grid prescriptions for the topology of the magnetic field (which affects the effective thermal conductivity), the magnitude and volume-filling factor of turbulence, which is calculated directly from the gravity/hydro solver (unlike previous work \citep{ruszkowski10,parrish10} in which volume-filling turbulence is inserted by hand), or turbulent heat diffusion (which is directly simulated). We have also simulated a cluster with significantly higher central density than in RO10, and still found it to be thermally stable. Other salient points include: 
\begin{itemize}
\item{In order for galaxies to excite volume-filling turbulence, rather than have turbulence confined to galactic wakes, they must excite $g$-modes, which requires that $\omega_{\rm stir} < \omega_{\rm BV}^{\rm MHD}$, where $\omega_{\rm BV}^{\rm MHD}$ is the \brunt frequency appropriate when thermal conduction timescales are rapid (equation \ref{eqn:brunt}). On the other hand, overwhelming the stabilizing buoyancy forces to randomize the magnetic field requires that $\omega_{\rm stir} > \omega_{\rm BV}$. These two requirements can be simultaneously satisfied since $\omega \propto l^{-2/3}$ for Kolomogorov turbulence; hence, the low-frequency, large scale modes can be trapped, while the high-frequency, small scale modes overcome the HBI.}
\item{We observed strong growth in vorticity, which is a good tracer of the growth of $g$-modes. We also observed turbulent amplification of B-fields in tandem with vorticity.}
\item{Thermal conduction provided about $\sim 30-50\%$ of the heating budget, with the rest coming from turbulent heat diffusion. Viscous dissipation of turbulent motions (and hence dynamical friction heating) is negligible. Turbulent heat diffusion tends to be more important in the center of the cluster, while conduction plays a greater role further out. The predominance of turbulent heat diffusion in the center---which is powered by motions on large scales---implies that it exhibits oscillations about the equilibrium temperature profile, and can occasionally exhibit small temperature inversions as high entropy fluid elements are compressed near the center. However, conduction plays a crucial part of the story; our most extreme stirring case still suffered a cooling catastrophe if thermal conduction was omitted. Besides supplying heat further out in the cluster, conduction also reduces stabilizing buoyancy forces and enables more efficient turbulent heat diffusion. It appears that turbulence enables conduction to operate, as well as vice-versa. The details of the interplay between turbulence and conduction, as well as the diffusion of metals in our stirring simulations, are interesting topics for future work.} 

\end{itemize} 

In this paper, we have focussed on a time-steady source of turbulence---stirring by galaxy motions---but we stress that other intermittent sources of turbulence such as mergers or AGN outbursts, can also contribute. Indeed, a sudden rise in heat transport processes such as conduction and turbulent heat diffusion due to an increase in turbulence could effect a CC to NCC transition \citep{guo09,ruszkowski10,parrish10}. Other processes which could reorient field lines in galaxy cluster include rising bubbles, which could amplify and straighten magnetic fields in their wake \citep{ruszkowski07,guo08b,bogdanovic09}. In the future, observations of Faraday rotation by SKA \citep{bogdanovic10} or magnetic draping around galaxies orbiting the cluster center \citep{pfrommer10}, could probe the topology of magnetic field lines and test these ideas. 
Finally, these ideas about the interplay between between the thermal conduction, the HBI and turbulence in the inner regions of the cluster also apply with equal force to the interplay between conduction, the MTI and turbulence in the outer regions of the cluster, which we present elsewhere (Ruszkowski et al 2010). 

\section*{Acknowledgments}
The software used in this work was in part developed by the DOE-supported ASC/Alliance Center for
Astrophysical Thermonuclear Flashes at the University of Chicago.
MR thanks Jeremy Hallum for his invaluable help with maintaining the computing cluster at the Michigan Academic 
Computing Center where most of the 
computations were performed. 
We are indebted to Dongwook Lee, the author of the MHD module in {\it FLASH} for letting us use a proprietary 
three-dimensional set of MHD modules. 
MR thanks Justin Nieusma for technical assistance with performing the simulations.
We thank Ian Parrish, Eliot Quataert, Elena Rasia, Prateek Sharma, Min-Su Shin, Maxim Markevitch, John ZuHone, 
Paul Nulsen, Aneta Siemiginowska, Christine Jones, Larry David, Bill Forman, Renato Dupke, Milos Milosavljevi{\'c},
John ZuHone, Matt Kunz, and Alex Schekochihin for discussions. 
SPO acknowledges support by NASA grant NNG06GH95G, and NSF grant 0908480. 
MR and SPO thank Institute of Astronomy, Cambridge, UK and Max Planck Institute for Astrophysics, Garching, 
Germany for their hospitality.

\bibliography{master_references} 

\label{lastpage}

\end{document}